\documentclass[longauth]{aa} 

\usepackage{natbib}
\bibpunct{(}{)}{;}{a}{}{,}

\usepackage{graphicx}
\usepackage{float}
\usepackage{txfonts}
\usepackage{xcolor}

\usepackage{hyperref}
\hypersetup{
    colorlinks=true,
    citecolor=blue,
    linkcolor=blue,
    filecolor=magenta,      
    urlcolor=cyan,
    pdftitle={Pyykkinen et al. 2025},
    pdfpagemode=FullScreen,
}

\newcommand{\orcid}[1]{\href{https://orcid.org/#1}{\includegraphics[width=10pt]{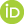}}}

\begin{document}

   \titlerunning{The H-free CS interaction in the Type Ib SN~2021efd}
   \authorrunning{N. Pyykkinen et al.}

   \title{The hydrogen-free circumstellar interaction in the Type Ib supernova 2021efd: A clue to the mechanism of the helium-layer stripping}

   \subtitle{}

\author{
N. Pyykkinen\inst{\ref{uni:Turku},\ref{NOT}}\orcid{0009-0006-0165-6986}
\and 
T.~Nagao\inst{\ref{uni:Turku},\ref{uni:aaltoradio},\ref{uni:aaltoengin},\ref{NAOJ}}\orcid{0000-0002-3933-7861}
\and  
H. Kuncarayakti\inst{\ref{uni:Turku},\ref{finca}}\orcid{0000-0002-1132-1366}
\and 
M. D. Stritzinger\inst{\ref{uni:aarhus}}\orcid{0000-0002-5571-1833}
\and 
T. Kangas\inst{\ref{uni:Turku},\ref{finca}}\orcid{0000-0002-5477-0217}
\and 
K. Maeda\inst{\ref{uni:kyoto}}\orcid{0000-0003-2611-7269}
\and
P. Chen \inst{\ref{chen1},\ref{chen2}}\orcid{0000-0003-0853-6427}
\and
J. Sollerman\inst{\ref{uni:klein}}\orcid{0000-0003-1546-6615}
\and
C. Burns\inst{\ref{uni:carnegie}}\orcid{0000-0003-4625-6629}
\and 
S. Bose\inst{\ref{uni:aarhus}}\orcid{0000-0003-3529-3854}
\and
G. Folatelli\inst{\ref{uni:argentina1},\ref{uni:argentina2},\ref{kavli}}\orcid{0000-0001-5247-1486}
\and
L. Ferrari\inst{\ref{uni:argentina1},\ref{uni:argentina2}}\orcid{0009-0000-6303-4169}
\and
N. Morrell \inst{\ref{carnegie:chile}}\orcid{0000-0003-2535-3091}
\and
A. Reguitti\inst{\ref{padova},\ref{merate}}\orcid{0000-0003-4254-2724}
\and
I. Salmaso\inst{\ref{salmaso}}\orcid{0000-0003-1450-0869}
\and
S. Mattila \inst{\ref{uni:Turku},\ref{cyprys}}\orcid{0000-0001-7497-2994}
\and
A. Gal-Yam \inst{\ref{weizmann}}\orcid{0000-0002-3653-5598}
\and
C. Fremling \inst{\ref{caltech} \orcid{0000-0002-4223-103X}}
\and
S. Anand \inst{\ref{stanford} \orcid{0000-0003-3768-7515}}
\and
M. M. Kasliwal \inst{\ref{caltech}}\orcid{0000-0002-5619-4938}
\and
C. P. Guti\'errez\inst{\ref{barcelona1},\ref{barcelona2}}\orcid{0000-0003-2375-2064}
\and
L. Galbany \inst{\ref{barcelona2},\ref{barcelona1}}\orcid{0000-0002-1296-6887}
\and
W. Hoogendam \inst{\ref{uni:hawaii}}\orcid{0000-0003-3953-9532}
\and
S. Schulze \inst{\ref{northwest}}\orcid{0000-0001-6797-1889}
\and
C. Ashall \inst{\ref{uni:hawaii}}\orcid{0000-0002-5221-7557}
\and
K. Medler \inst{\ref{uni:hawaii}}\orcid{0000-0001-7186-105X}
\and
C. M. Pfeffer \inst{\ref{uni:hawaii}}\orcid{0000-0002-7305-8321}
\and
P. Lundqvist\inst{\ref{uni:klein}}\orcid{0000-0003-0065-2933}
\and
B. Rusholme \inst{\ref{caltech2}}\orcid{0000-0001-7648-4142}
\and
J. Adler \inst{\ref{caltech2}}\orcid{0009-0006-7265-2747}
}
 \institute{
            Department of Physics and Astronomy, University of Turku, FI-20014 Turku, Finland\label{uni:Turku}
            \and
            Nordic Optical Telescope, Rambla José Ana Fernández, Pérez 7, E-38711 Breña Baja, Spain \label{NOT}
            \and
            Aalto University Mets\"ahovi Radio Observatory, Mets\"ahovintie 114, 02540 Kylm\"al\"a, Finland\label{uni:aaltoradio}
            \and
            Aalto University Department of Electronics and Nanoengineering, P.O. BOX 15500, FI-00076 AALTO, Finland\label{uni:aaltoengin}
            \and
            National Astronomical Observatory of Japan, National Institutes of Natural Sciences, 2-21-1 Osawa, Mitaka, Tokyo 181-8588, Japan \label{NAOJ}
            \and
            Finnish Center of Astronomy with ESO (FINCA), 20014 University of Turku, Vesilinnantie 5, Turku, Finland \label{finca}
            \and 
            Department of Physics and Astronomy, Aarhus University, Ny Munkegade 120, DK-8000 Aarhus C, Denmark Aarhus University\label{uni:aarhus}
            \and
            Department of Astronomy, Kyoto University, Kitashirakawa-Oiwake-cho, Sakyo-ku, Kyoto, 606-8502, Japan\label{uni:kyoto}
            \and
            Institute for Advanced Study in Physics, Zhejiang University, Hangzhou 310027, China \label{chen1}
            \and
            Institute for Astronomy, School of Physics, Zhejiang University, Hangzhou 310027, China \label{chen2}
            \and
            The Oskar Klein Centre, Department of Astronomy, Stockholm University, AlbaNova, SE-10691, Stockholm, Sweden \label{uni:klein}
            \and
            Observatories of the Carnegie Institution for Science, 813 Santa Barbara St., Pasadena, CA 91101, USA\label{uni:carnegie}
            \and
            Instituto de Astrofísica de La Plata (UNLP - CONICET), Paseo del Bosque S/N, 1900 La Plata, Argentina \label{uni:argentina1}
            \and
            Facultad de Ciencias Astronómicas y Geofísicas (UNLP), Paseo del Bosque S/N, 1900 La Plata, Argentina \label{uni:argentina2}
            \and
            Kavli Institute for the Physics and Mathematics of the Universe (WPI),The University of Tokyo
            Institutes for Advanced Study, The University of Tokyo, Kashiwa, Chiba 277-8583, Japan \label{kavli}
            \and
            Carnegie Observatories, Las Campanas Observatory, Casilla 601, La Serena, Chile \label{carnegie:chile}
            \and
            INAF – Osservatorio Astronomico di Padova, Vicolo dell'Osservatorio 5, I-35122 Padova, Italy \label{padova}
            \and
            INAF – Osservatorio Astronomico di Brera, Via E. Bianchi 46, I-23807 Merate (LC), Italy \label{merate}
            \and 
            INAF–Osservatorio Astronomico di Capodimonte, Salita Moiariello 16, 80131 Napoli, Italy\label{salmaso}
            \and
            School of Sciences, European University Cyprus, Diogenes Street, Engomi, 1516 Nicosia, Cyprus\label{cyprys}
            \and
            Department of Particle Physics and Astrophysics Weizmann Institute of Science 234 Herzl St. Rehovot, Israel \label{weizmann}
            \and 
            Division of Physics, Mathematics, and Astronomy, California Institute of Technology, Pasadena, CA 91125, USA \label{caltech}
            \and
            Kavli Institute for Particle Astrophysics and Cosmology, Stanford University, Stanford, CA94305-4085, USA \label{stanford}
            \and
            Institut d'Estudis Espacials de Catalunya (IEEC), Edifici RDIT,
            Campus UPC, 08860 Castelldefels (Barcelona), Spain\label{barcelona1}
            \and
            Institute of Space Sciences (ICE, CSIC), Campus UAB, Carrer de Can Magrans, s/n, E-08193 Barcelona, Spain \label{barcelona2}
            \and
            Institute for Astronomy, University of Hawai'i at Manoa, 2680 Woodlawn Dr., Hawai'i, HI 96822, USA \label{uni:hawaii}
            \and
            Center for Interdisciplinary Exploration and Research in Astrophysics (CIERA), Northwestern University, 1800 Sherman Ave., Evanston, IL 60201, USA \label{northwest}
            \and
            IPAC, California Institute of Technology, 1200 E. California Blvd, Pasadena, CA 91125, USA \label{caltech2}
            }

   \date{Received ***; accepted ***}

  \abstract 
   {Stripped-envelope supernovae (SESNe), including Type~IIb, Ib, and Ic supernovae (SNe), originate from the explosions of massive stars whose outer envelopes have been largely removed during their lifetimes. The main stripping mechanism for the hydrogen (H) envelope in the progenitors of SESNe is often considered to be interaction with a binary companion, while that for the helium (He) layer is unclear.}
   {We aim to study the process of the He-layer stripping in the progenitors of SESNe, which is closely related to the origin of their diverse observational properties.}
   {We conducted photometric and spectroscopic observations of the Type~Ib SN~2021efd, which shows signs of interaction with H-free circumstellar material (CSM). At early phases, its photometric and spectroscopic properties resemble those of typical Type~Ib SNe. Around 30 days after the $r$-band light curve (LC) peak, until at least $\sim 770$ days, the multi-band LCs display excessive luminosity compared to regular SESNe and at least three distinct peaks. The light curve evolution is similar to that of SN~2019tsf, whose previously unpublished spectrum at 400 days is also presented here.
   The nebular spectrum of SN~2021efd shows narrow emission lines ($\sim1000$ km s$^{-1}$) in various species, such as 
   \ion{O}{i}, \ion{Ca}{ii}, \ion{Mg}{ii}, \ion{He}{i}, [\ion{O}{i}], [\ion{Ca}{ii}], and [\ion{S}{ii}]. Based on the observations, we studied the properties of the ejecta and CSM of SN~2021efd.
   }
   {Our observations suggest that SN~2021efd is a Type Ib SN interacting with CSM with the following parameters: The estimated ejecta mass, explosion energy, and $^{56}$Ni mass are 2.2 $M_\odot$, $9.1\times10^{50}~\mathrm{erg}$, of 0.14 $M_\odot$, respectively, while the estimated CSM mass, composition, and distribution are at least a few times 0.1 M$_{\odot}$, H-free, and clumpy, respectively.
   Based on the estimated ejecta properties, we conclude that this event is a transitional SN whose progenitor was experiencing He-layer stripping at the epoch of the explosion, and was on the way to becoming a carbon-oxygen star (as the progenitors of Type~Ic SNe) from a He star (as the progenitors of Type Ib SNe). 
   The estimated CSM properties suggest that the progenitor had some episodic mass ejections with the rate of $\sim 5\times 10^{-3}-10^{-2}~M_\odot~\mathrm{yr}^{-1}$ for the last decade and slightly smaller before this final phase at least from $\sim200$ years before the explosion, for the assumed CSM velocity of 100 km s$^{-1}$. For the case of $\sim 1000$ km s$^{-1}$, the necessary mass-loss rate would be increased by a factor of ten, and the timescales decreased by a factor of ten.
   }
   {}


\maketitle

\section{Introduction}

Massive stars ($M \gtrsim 8 ~M_\odot$) terminate their lives in catastrophic explosions known as core-collapse supernovae (CCSNe). CCSNe are classified into several types based on their observational properties. CCSNe containing hydrogen (H) lines in their spectra are categorized as Type~II SNe, while those with little to no H in their spectra are collectively called stripped-envelope supernovae (SESNe), which include the Type IIb, Ib, and Ic \citep[e.g.,][]{filippenko97,gal-yam17}.
 SNe~Ib show helium (He) lines in their spectra around peak brightness, while SNe~Ic do not exhibit H nor He lines. The spectroscopic properties of  SNe~IIb are similar to those of SNe~II at early phases, but after the light curve maximum, their spectra change, becoming identical to those of Type Ib SNe.

The observational diversity in CCSNe is considered to reflect the different mass-loss histories of their progenitors before the explosions \citep[e.g.,][]{nomoto95, gal-yam17, dessart24}. If the H-envelope of a progenitor star is almost entirely stripped by the mass loss before the terminal explosion, it would explode as a Type~Ib SN. If a star also loses a majority of its He-layer, it explodes as a Type~Ic SN. A star with the H-envelope largely intact would be a Type~II SN, while a star with a mostly stripped H-envelope would produce a Type~IIb SN.

The origin of the difference in the observational properties and thus in the mass-loss history in CCSNe has been discussed over the last decades. The initial mass and metallicity of massive stars are important parameters that govern their evolution. For estimating the initial masses of the progenitors of CCSNe, a method based on the [\ion{O}{i}]/[\ion{Ca}{ii}] ratio in the nebular spectra is often used, where this ratio is correlated with the carbon-oxygen core mass of the progenitor and thus its zero-age main sequence (ZAMS) mass \citep[e.g.,][]{fransson89, kuncarayakti15, jerkstrand17, dessart21b, dessart23}. Based on a statistical analysis of the nebular spectra of SESNe, \citet{fang19} and \citet{fang22} concluded that the initial masses of the progenitors of Type IIb and Ib are similar and systematically lower than those of Type~Ic SNe. This conclusion is also supported by environmental studies \citep[e.g.,][]{anderson12,kangas17, kuncarayakti18b}.
Additionally, \citet{barmentloo24} also investigated the progenitor masses of SESNe using a new method based on the [N II] $\lambda\lambda$6548, 6583 lines in their nebular spectra, and also concluded that Type~IIb SNe mostly come from stars with a low He core mass ($\sim 3$ M$_{\odot}$), whose ZAMS masses are similar to the progenitors of typical Type~II SNe roughly 8-17 $M_\odot$ \citep[e.g.,][]{smartt09}. We note that certain, mostly peculiar, Type IIb and Ib SNe arise from progenitors with higher initial masses than those considered here \citep[e.g.,][]{stritzinger2020,schweyer2025}. Based on the observational conclusion that the progenitors of Type~IIb and Ib SNe originate from similarly low-mass massive stars, \citet{fang19} concluded that the H-envelope stripping is a mass-insensitive process. A promising candidate is interaction with a binary companion \citep[e.g.,][]{nomoto95, maund04, smith11, dessart11, bersten12, eldridge13, yoon17}. This scenario explains well, not only observational properties \citep[e.g.,][]{dessart24}, but also the observational relations between the progenitor radius and mass-loss rate in Type IIb SNe \citep{ouchi17}.
On the other hand, since the progenitors of Type~Ic SNe are on average more massive than those of Type~IIb and Type~Ib SNe, the He-layer stripping is possibly a mass-dependent process \citep{fang19}. In addition, the environments of Type~Ic SNe have higher metallicities than those of Type IIb and Ib SNe \citep[e.g.,][]{anderson15,kuncarayakti18b}, implying that the mechanism for the He-layer stripping might also depend on the metallicity.
While simulations support binary stripping for H-envelope removal in SESN progenitors, they disfavor it for He-layer stripping in typical Type Ic progenitors, as such extreme stripping is difficult to reproduce through binary interactions \citep[e.g.,][]{yoon17,dessart24}. Strong stellar winds \citep [e.g.,][]{crowther07} and pre-SN eruptions \citep [e.g.,][]{crowther07,fuller18,wu21} are other possibilities. The detailed nature of this process is theoretically and observationally unclear.

In some rare SESNe, significant interactions between the SN ejecta and their circumstellar material (CSM) have been observed. Since the origins of the CSM are due to mass loss from the progenitor systems before the explosion, interacting SESNe provide good opportunities to unravel the mass-loss histories of SESN progenitor systems. 
Types Ibn and Icn SNe are characterized by interaction with He-rich and C/O-rich CSM, respectively \citep[e.g.,][]{pastorello07,gal-yam22, pellegrino22}. They display narrow lines in their early spectra, which are the result of this interaction. Their spectra differ from regular Type Ib and Ic SNe beyond just the presence of narrow lines, and they generally show faster light curve evolution and higher peak brightness than regular Type~Ib and Type~Ic SESNe. This suggests that the progenitor systems of Ibn SNe and SESNe might not be related.
Binary interaction and strong winds or outbursts have been suggested as the mass-loss mechanism for Type~Ibn/Icn SN progenitors \citep{hosseinzadeh17,maeda22}.
Some Type~Ibn and Type~Icn SNe have been suggested to be the explosions of Wolf-Rayet (WR) stars more massive than regular SESNe progenitors \citep[e.g.,][]{pastorello07,tominaga08}. \citet{gal-yam22}, suggesting that the differences between Type~Ibn and Type~Icn SNe mirror those of WN and WC subclasses of WR stars. Other scenarios include explosions of relatively low-mass stars surrounded by massive CSM \citep[e.g.,][]{moriya16, dessart22}, and mergers between a WR star and a compact companion \citep{metzger22}. Recently, \citet{schulze25} suggested that SN~2021yfj resulted from the explosion of an ultra-stripped star, classified as a Type~Ien SN, interacting with sulfur/silicon-rich CSM.

There are also some other interacting SESNe whose photometric and/or spectroscopic properties are similar to those of typical SESNe for some time, but with some signs of extensive circumstellar interaction (CSI). For example, the presence of a blue continuum, narrow lines in their spectra, and/or slow photometric evolution.
Some are colliding with H-rich CSM, such as, Type Ib SN 2019oys \citep{sollerman20} and Type Ib SN~2019yvr \citep{ferrari24}.
In one case, a He-poor Type~Ic SN~2017dio evolved into a Type IIn-like SN with narrow H lines \citep{kuncarayakti18}. Since the outer layers of the progenitors of Type Ic SNe are H-poor, the observed H-rich CSM should originate from the companion star in the binary system rather than from the progenitor itself. Hence, the CSM in systems like this preserves information on mass loss from periods of intense binary interaction.
In some cases, SESNe are thought to be interacting with H-poor CSM. The Type~Ib SN~2019tsf \citep{sollerman20,zenati22} is one of these cases, where it exhibits a bumpy light curve (LC) with a slow decline. 
The broad-lined Type Ic SN~2022xxf is another object with a bumpy LC \citep{kuncarayakti23}, suggesting that it may have been interacting with H/He-poor CSM. The Type~Ic SNe~2010mb \citep{ben-ami14} and 2021ocs \citep{kuncarayakti22} are also inferred to be SNe interacting with H/He-poor CSM. 
In such cases, where H-poor CSM likely results from envelope stripping in progenitor stars rather than mass-loss of the companion star, it is essential to characterize the CSM.

\begin{figure}
    \centering
    \includegraphics[width=\linewidth]{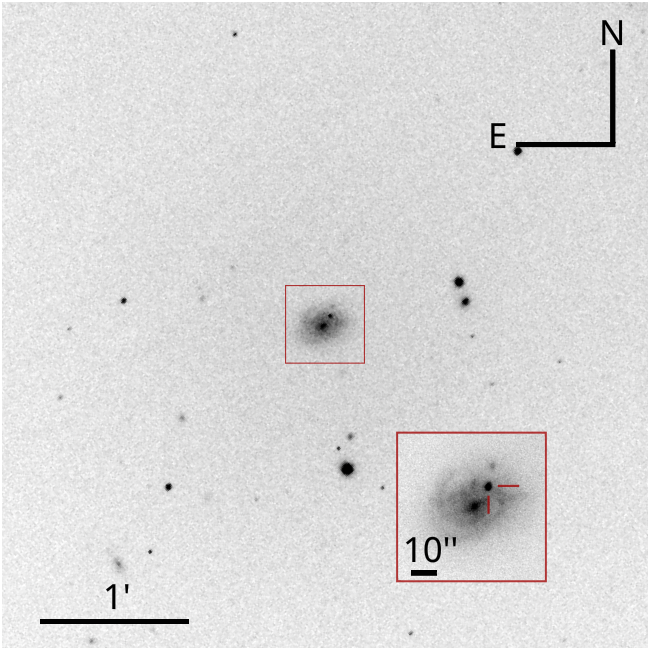}
    \caption{Acquisition image of SN~2021efd and its host galaxy, taken with the NOT using ALFOSC without a filter on 4th April 2021 UT (MJD 59318.94). The red markers in the zoomed-in image mark the location of SN~2021efd in the host galaxy.}
    \label{fig:acq}
\end{figure}

In this article, we present observations of the peculiar Type~Ib SN~2021efd with multiple peaks in its light-curve evolution, indicating strong interaction with CSM. SN~2021efd was discovered on 2.79 March 2021 UT (i.e., MJD $= 59275.34$) by the Zwicky Transient Facility (ZTF) using the 1.2-m Palomar Oschin telescope \citep{masci19,bellm19, graham19,dekany2020}. The last non-detection prior to the discovery is based on an \textit{r}-band image taken by ZTF on MJD 59269.27 with a limiting magnitude of >21.35 mag. We estimate the explosion date in Sect.~\ref{sec: explosion date estimate}. All phases are in the rest-frame and are henceforth displayed relative to the epoch of the $r$-band maximum (MJD 59289.15 $\pm 0.99)$. 

The host galaxy of SN~2021efd is KUG~1121+239 (see Fig.~\ref{fig:acq}), which has a spectroscopic redshift of $z= 0.027756\pm 0.000016$ \citep{sdss17}. We adopt the luminosity distance of $116.4~\mathrm{Mpc}$ inferred from this redshift with the assumption of a flat cosmology and Hubble constant ($H_0$) of $73~\mathrm{km}~\mathrm{s}^{-1}\mathrm{Mpc}^{-1}$, $\Omega_\mathrm{m}=0.3$ and $\Omega_\Lambda=0.7$. Based on a spectrum obtained with the 2.56-m Nordic Optical Telescope (NOT) 1.81 days after discovery,  \cite{reguitti21} classified SN~2021efd as a SN~Ibc. 
Our analysis shows the spectra to be more similar to  SNe~Ib than to Type Ic SNe (see Sect.~\ref{sec:spec_properties}).
Due to the lack of significant Na~{\sc I}~D absorption lines in the SN spectrum at the redshift of the host galaxy, we assume the extinction within the host galaxy to be minimal, and thus correct the photomeric data only for the Milky-Way (MW) extinction of $E(B-V)_{MW} =0.0168$ mag \citep{schafly11}.

The organization of this paper is as follows. Sect.~\ref{sec:obs} presents the observations. Photometric and spectroscopic properties are detailed in Sects.~\ref{sec:phot_properties} and \ref{sec:spec_properties}, respectively. Sect.~\ref{sec:SN_properties} estimates the key properties of SN~2021efd. The origin of the brightness excess is examined in Sect.~\ref{sec:origin}, along with the derivation of the CSM properties through LC modeling. Sect.~\ref{sec:discussion} addresses the mass-loss mechanism responsible for creating the CSM in SN~2021efd and its progenitor. The paper concludes with a summary in Sect.~\ref{sec:conclusion}.      

\section{Observations and data reduction}\label{sec:obs}
\subsection{Photometry}

\begin{table*}[t]
\caption{Log of spectral observations of SN 2021efd.\centering}
\small
\begin{tabular}{c c c c c c c c} \hline
Phase\tablefootmark{a} (days) & Date (UT)\tablefootmark{b} & MJD\tablefootmark{b} & Telescope & Instrument & Grism & Range (\AA) & Resolving power \\
\hline
$-12$ & 04 Mar.2021 & 59277.15 & NOT & ALFOSC & Gr\#4 & 3200-9600 & 360  \\
$-$8 & 04 Mar.2021 & 	59281.23 & Baade & IMACS & Gri-300-17.5 & 3900-8000 & 1240  \\
$-$1 & 15 Mar.2021 & 59288.10 & NOT & ALFOSC & Gr\#4 & 3200-9600 & 360  \\
+17 & 02 Apr.2021 & 59306.94 & NOT & ALFOSC & Gr\#4 & 3200-9600 & 360  \\
+29 & 14 Apr.2021 & 59318.94 & NOT & ALFOSC & Gr\#4 & 3200-9600 & 360  \\
+315 &03 Feb.2022 &	59613.00 & Keck & LRIS &400/3400-400/8500 & 3100-10300 & 1000
\\
+322 & 10 Feb.2022 & 59620.22 & VLT & FORS2 & GRIS 300V & 4450-8700 & 440  \\
+374 &04 Apr.2022 &	59673.54 & Keck & LRIS &400/3400-400/8500 & 3100-10300 & 1000
\\
+396 &27 Apr.2022 &59696.00 & Keck & LRIS &600/4000-1200/7500 & 3100-7100 & -
\\
+460 &02 Jul.2022 &59762.28 & Keck & LRIS &400/3400-400/8500 & 3100-10300 & 1000
\\
+624 &17 Dec..2022 &59930.46 & MMT & Binospec & Grism 270 & 3900-9200 & 1340
\\
+741 &16 Apr.2023 &60050.53 & Keck & LRIS &400/3400-400/8500 & 3100-10300 & 1000
\\
+750 &26 Apr.2023 &60060.00 & Keck & LRIS &400/3400-400/8500 & 3100-10300 & 1000\\
\hline 
\end{tabular}
\tablefoot{\\
\tablefoottext{a}{Days relative to the epoch of $r$-band maximum in rest-frame}.\tablefoottext{b}{Days in observer frame.} 
}
\normalsize
\label{tab:Spectra_log}
\end{table*}

We present 40 epochs of multi-band ($uBVgri$) photometry of SN~2021efd obtained with the Las Campanas Observatory's (LCO) 1-m Henrietta Swope telescope equipped with a direct imaging camera. These data were obtained as part of the Precision Observations of Infant Supernova Explosions (POISE) collaboration \citep{burns21} and extend from $-$12.6 to $+$27.2 days with three additional epochs between +390.6 and +392.1 days. Imaging data were reduced following standard procedures \citep{Hamuy2006}, and 
the instrumental magnitudes from the point-spread-function (PSF) photometry of the SN were calibrated to apparent magnitudes using nightly zero-points inferred from a local sequence of stars calibrated by standard star observations. The host galaxy flux was subtracted with a template photometry taken three years after the explosion.

We also present eight epochs of late-time $gri$ photometry from the NOT using the Alhambra Faint Object Spectrograph and Camera (ALFOSC), two epochs of $r$ photometry with the Palomar 60-inch telescope (P60) using SED Machine (SEDM, \citealt{nadia2018}), and one epoch of $griz$ photometry with the 6.5-m Multi-Mirror Telescope (MMT) using the Binospec instrument.
These data were obtained as part of the Zwicky Transient Facility collaboration. Photometry of ALFOSC and Binospec data was performed following the procedures as described in \cite{Chen2022}. We used archival images from the Pan-STARRS1 (PS1) survey as reference images for image subtraction. The instrumental magnitudes were obtained with PSF photometry on the subtracted images. The photometric zero-point calibration was done using the reference stars in the field view of the science image.

Our photometry of SN~2021efd is complemented with measurements obtained by the ZTF survey (\citealt{bellm19, graham19, masci19}) and the ATLAS Project (Asteroid Terrestrial-impact Last Alert System;  \citealt{tonry18, smith20}). Photometry from these public surveys were retrieved through the ZTF Forced Photometry Service$^1$\footnotetext[1]{\url{https://ztfweb.ipac.caltech.edu/cgi-bin/requestForcedPhotometry.cgi}} \citep{masci2023newforcedphotometryservice} and the ATLAS Forced Photometry Service \citep{Shingles2021}$^2$\footnotetext[2]{\url{Fallingstar-data.com/forcedphot/}}, respectively. These photometric measurements were obtained on science images having undergone host-galaxy template subtraction. We adopted a signal-to-noise ratio of 3 as the detection limit for the forced photometry. The photometry from these surveys was binned into one-day intervals.

\subsection{Spectroscopy}

We present 13 epochs of optical spectroscopy of SN~2021efd extending from $-12$ days to +771~days. Four of these were obtained with the NOT equipped with ALFOSC as part of the NOT Unbiased Transient Survey (NUTS2) collaboration. A pre-maximum spectrum was also obtained by POISE with the 6.5-m Magellan Baade telescope equipped with the Inamori-Magellan Areal Camera and Spectrograph (IMACS). One nebular spectrum was obtained with the Very Large Telescope (VLT) equipped with the FOcal Reducer and low dispersion Spectrograph 2 (FORS2). Additionally, six nebular spectra were obtained by the ZTF collaboration with the Keck telescope using the Low Resolution Imaging Spectrograph (LRIS), and one with the MMT using the Binospec instrument. The nebular spectra span from +324 days until +771 days.
A journal of spectroscopic observations is provided in Table~\ref{tab:Spectra_log}.

Spectroscopic data were reduced following standard procedures, including bias subtraction, flat-field correction, spectral extraction, wavelength correction, and flux calibration relative to sensitivity functions derived from observations of spectroscopic flux standard stars. 
We reduced the spectroscopic data using the \texttt{ALFOSCGUI}$^3$\footnotetext[3]{\url{https://sngroup.oapd.inaf.it/foscgui.html}} and ESO pipelines for the ALFOSC and FORS spectra, respectively. 
The IMACS spectrum was reduced with standard IRAF procedures. For the Binospec spectrum, the basic data processing (bias subtraction, flat fielding) is done using the Binospec pipeline \citep{Kansky2019}. The spectrum is reduced with \texttt{IRAF}, including cosmic-ray removal, wavelength calibration (using arc lamp frames taken immediately after the target observation), and relative flux calibration with archived spectroscopic standards observation. The LRIS spectra \citep{oke95} were reduced with {\tt Lpipe} \citep{Perley19}. 
The nebular phase spectra and the IMACS spectrum were also corrected for telluric features. In the case of the IMACS spectrum, residuals of the telluric feature remain.  

We obtained the nebular spectrum of SN~2019tsf at +418 days under the same program and setup as the FORS2 spectrum of SN~2021efd, except with an order blocking filter. This phase was not covered by the previous studies of SN~2019tsf \citep{sollerman20, zenati22}. Spectroscopic data are available for download in electronic format on WISeREP (Weizmann Interactive Supernova Data Repository )\footnotetext[4]{\url{https://www.wiserep.org/}}.

\section{Photometric properties} \label{sec:phot_properties}

\subsection{Explosion date estimate} \label{sec: explosion date estimate}

We estimate the explosion date by fitting the early phases of the LC with a power-law curve of $f_0(t-t_0)^n$. Here, $t_0$ is the explosion date, $f_0$ is a scaling coefficient in flux units, and $n$ is the power-law index. The  $r$-band data for this analysis were used because of the comprehensive coverage of the early phases. The epochs chosen for the fitting are from $-11$  to $-5$~days before the peak. 
We estimate the explosion date to be MJD $\approx$59273.7$\pm$1.2. This is 15.5$\pm$1.2 days prior to the observed $r$-band peak. The last non-detection prior to discovery, obtained on MJD 59269.27 with a ZTF $r$-band limiting magnitude of $>21.35$, occurred $\sim4.4$ days before this estimated explosion date.

\subsection{Light curve evolution}

\begin{figure*}[ht]
\centering
\includegraphics[width=0.90\linewidth]{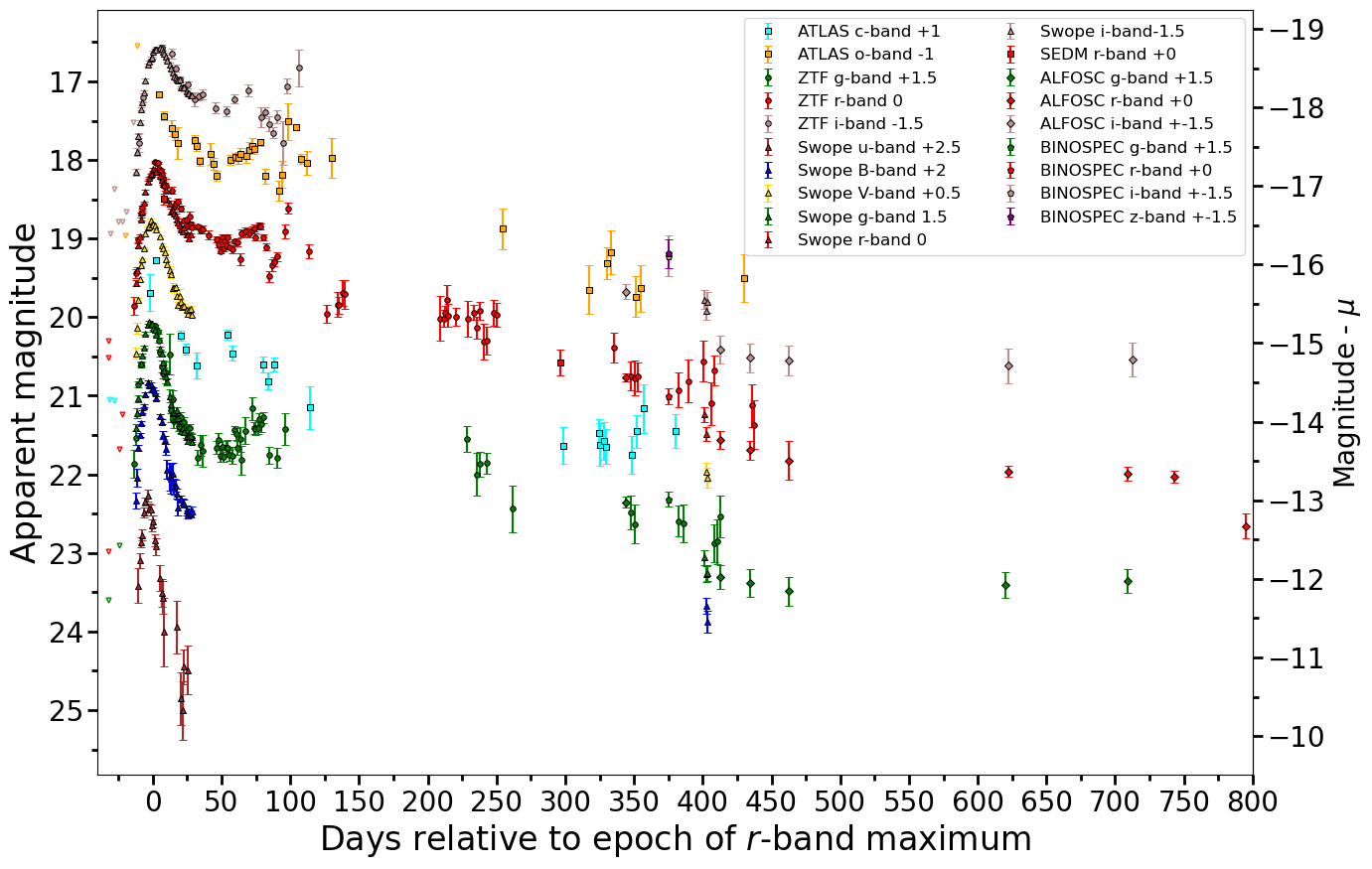}
\caption{Optical LCs of SN~2021efd  corrected for MW extinction. Photometry obtained by POISE, ATLAS, and ZTF is plotted with triangles, squares, and circles, respectively. Unfilled triangles correspond to ZTF and ATLAS non-detections obtained prior to the first detections in each band.} 
\label{lc}   
\end{figure*}

\begin{figure} 
       \centering
        \includegraphics[width=\hsize]{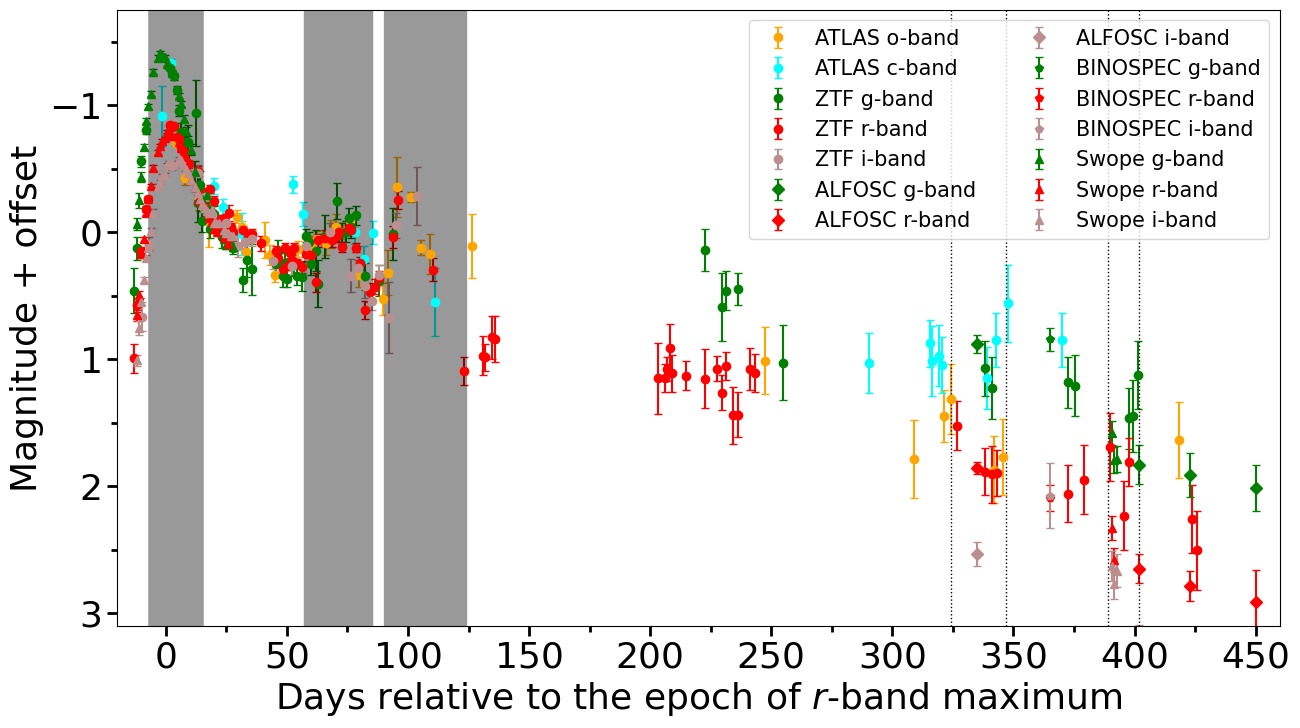}
        \caption{LCs of SN~2021efd, with magnitudes shifted so that the second peaks roughly match. The shaded areas highlight the clear bumps at $+0$ days, $\sim +75$ days, $\sim +105$ days. Dotted lines mark the possible later bumps at $\sim +325$ days, $\sim +345$ days, $\sim +390$ days, and $\sim +400$ days.}
        \label{bumps}
\end{figure}

The multi-band LCs of SN~2021efd are displayed in Fig.~\ref{lc}. 
The peak absolute $g$- and $r$-band magnitudes are $M_g =-16.77\pm 0.02~\mathrm{mag}$ and $M_r=-17.31\pm 0.01~\mathrm{mag}$, where the error is the  photometric error. The absolute magnitudes here are corrected for MW extinction, but no K-correction was applied. The late-time ($\gtrsim300$ days) photometric data points from the ZTF and ATLAS surveys are near the detection limits of these surveys, which is reflected by their large uncertainties. After the first peak, the LCs show rapid declines, which slow down at $\sim 30$ days from the LC peak, as well as a rebrightening starting from $\sim$ 60 days. 
The LCs of SN~2021efd exhibit at least three clear peaks at +0, +75, and +105 days; these are visible in Fig.~\ref{bumps}, where the LCs of all bands are overlaid.
The magnitudes in this figure are shifted to match the second peak of the LCs at 75 days. The durations of these bumps are roughly 30 days; the bump at $\sim 105$ days is more luminous than the one at $\sim 75$ days. The late phases of the LC of SN~2021efd appear to have more bumpy features at $\sim$ 325 days, $\sim$ 345 days, $\sim 390$ days, and $\sim$ 400 days that can not be clearly distinguished due to a low cadence of observations, and the large uncertainties.
The strengths and durations of the latter bumps after $\sim$ 300 days can not be constrained with certainty since there is a lack of detailed observations during the possible bump and around it. 

\begin{figure}[t]
        \centering
        \includegraphics[width=\hsize]{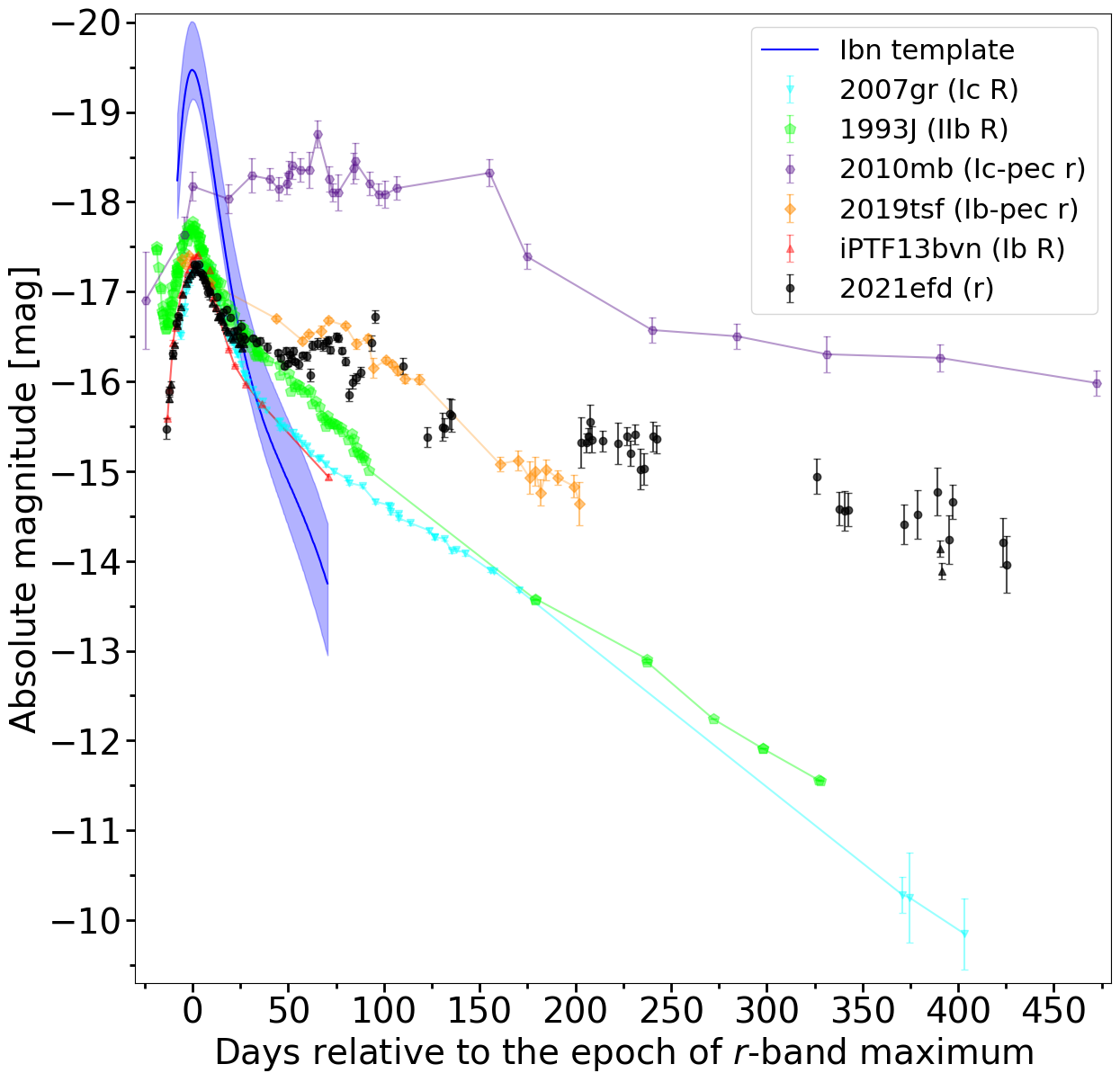}
        \caption{The LC of SN~2021efd compared to those of other SESNe and templates of Type~Ibn SNe from \cite{hosseinzadeh17}.
        The comparison SNe include a regular Type Ib SN iPTF13bvn \citep{srivastav14}, a regular Type~IIb SN~1993J \citep{richmond94}, a regular Type~Ic SN 2007gr \citep{hunter09}, and a peculiar Type~Ib SN~2019tsf \citep{masci19}.}
        \label{comp}
    \end{figure}

We compare the $r$-band LC of SN~2021efd to those of other well-observed SESNe including the Type~IIb SN~1993J \citep{richmond94}, the Type~Ib iPTF13bvn  \citep{srivastav14}, and the Type~Ic SN~2007gr \citep{hunter09} in Fig.~\ref{comp}. All of the LCs were corrected for extinction, which were determined in their respective articles. 
The first peak is similar to those of other SESNe, which are powered by the energy input from $^{56}$Ni. The rise time of SN~2021efd in the $r$~band to the peak is 15.5 days (see Sect.~\ref{sec: explosion date estimate}), which is intermediate between the typical values for Type~Ib and Type~Ic SNe, i.e., slightly shorter than those for Type Ib and longer than those for Type Ic SNe \citep[see Fig. 10 of][]{taddia15}. 
The decline rate after the first peak ($\Delta m_{15}\sim$ 0.57 mag for SN~2021efd from the POISE LC) is similar to the comparison SNe (0.83, 0.98, and 0.61 mag for SN~1993J, iPTF13bvn, and SN~2007gr, respectively). After $\sim$ 30 days from the $r$-band peak, its decline becomes slower, as if it enters the tail phase. 
The difference between the magnitudes at the peak and the beginning of the tail phase is smaller ($\Delta m_{30}\sim 0.86$ mag for SN~2021efd) than those for the other SNe (1.37, 1.58, and 1.41 mag for SN~1993J, iPTF13bvn, and SN~2007gr, respectively). This may imply that an additional powering mechanism starts to contribute to the LC evolution around this time, in addition to the radioactive decay power by $^{56}$Ni and $^{56}$Co (see Sect.~\ref{sec:origin}). 

The $r$-band LC of SN~2021efd at late times bears no resemblance to regular SESNe. Around 30 days after the $r$-band peak, the LC deviates from the linear decline, and at around 60 days, the light curve begins brightening again with two additional peaks at $\sim 75$ and 105 days. At the beginning of the tail phase during the period from 20 to 50 days, the decline rate is 1.49 mag per 100 days, which is roughly consistent with the 1.33 mag per 100 days expected from radioactive decay of both $^{56}$Ni and $^{56}$Co with full $\gamma$-ray trapping \citep{arnett1980,arnett1982}. The decline rate at the tail phase is 0.60 mag per 100 days during the period from 50 to 200 days from the $r$-band peak. This value is less than that expected from $^{56}$Co decay, 0.98 mag per 100 days in the case of the full $\gamma$-ray trapping \citep{arnett1980,arnett1982}.
SN~2019tsf, which has been interpreted to be a Type~Ib SN interacting with H-poor CSM \citep[e.g.,][]{sollerman20, zenati22}, has a similar bumpy LC. Its LC has a faster decay rate than that of SN~2021efd.
SN 2010mb, which is a Type~Ic SNe interacting with H/He-poor CSM \citep{ben-ami14}, exhibits a slower decline and brightness excess at earlier epochs compared to SN~2021efd.
The averaged $R$-band LC of Type~Ibn SNe has a much higher peak luminosity and a significantly higher decline rate than those of regular SESNe and SN~2021efd.

\subsection{Color evolution}
\begin{figure} [t]
        \begin{center}
        \includegraphics[width=\hsize]{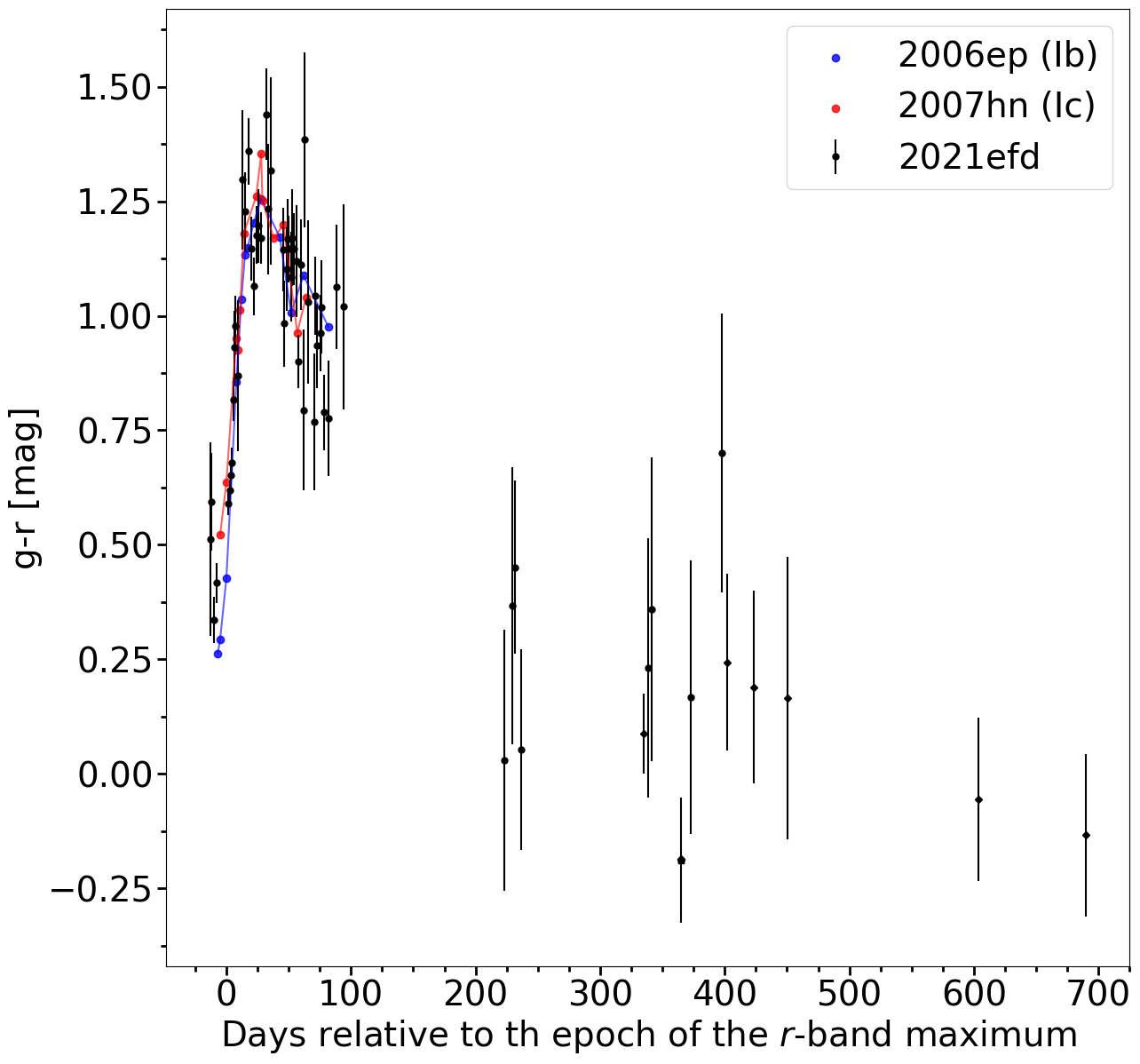}
        \caption{\textit{g-r} color evolution. The black dots show the evolution of SN~2021efd. The circles, diamonds, and pentagons correspond to ZTF, ALFOSC, and Binospec data, respectively. The blue line represents the color evolution of the Type Ib SN 2006ep, and the red line represents the Type Ic 2007hn presented in \citet{stritzinger2017a} }
        \label{color_comp}
        \end{center}
\end{figure}

The color evolution of SN~2021efd is presented in Fig.~\ref{color_comp}. 
We took all the ZTF photometry epochs that have both $g$- and $r$-band observations within one day and interpolated the magnitudes at these epochs. Using these interpolated magnitudes, we derive the $g-r$ color of SN 2021efd. At late times, colors from ALFOSC, Swope, and Binospec are also included. Every individual ($g-r$) color is derived using $g$ and $r$ photometry from the same instrument. The photometric data were corrected for MW extinction but not for that of the host galaxy. The color evolution of SN 2021efd around the first peak is similar to that of SN 2006ep and SN 2007hn, respectively a Type Ib and a Type Ic. During the rebrightenings, the color becomes bluer at a similar rate, although this is uncertain due to the noisy color evolution observed in SN 2021efd. At $\sim 100$ days, the uncertainties in the photometry of SN~2021efd become large, and the observation epochs on different bands do not always coincide. Once SN 2021efd re-emerged from behind the Sun at $\sim 200$ days, it had evolved to become much bluer. We note that, since the values of the magnitudes at late phases should be dominated by the strong spectral emission lines, the interpretation of the color evolution is not straightforward.

\subsection{Pseudo-bolometric light curve}

\begin{figure}
    \centering
    \includegraphics[width=\hsize]{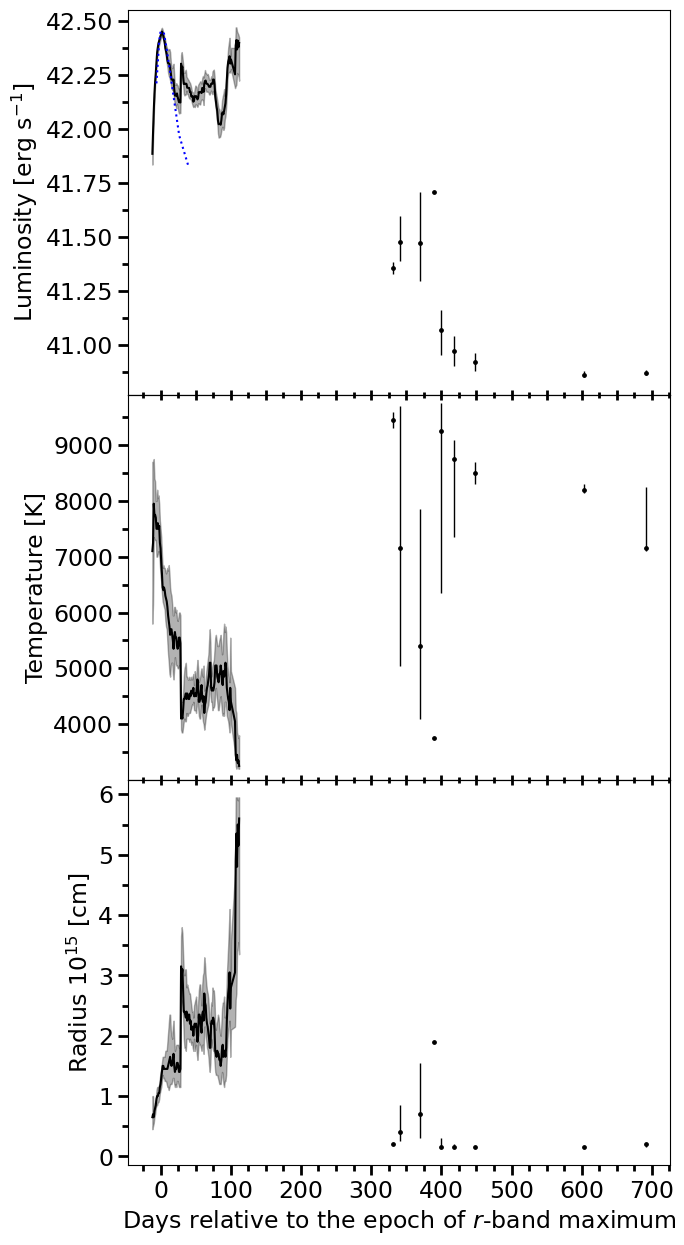}
    \caption{Results of the BB fitting. Top: The bolometric LC of SN~2021efd (black) and the bolometric luminosity of Type Ib SN 2004gv (blue) \citep{taddia18} indicated by a blue dotted line. Middle: The BB temperature of the photosphere. Bottom: The BB radius of the photosphere. The errors are the 1-$\sigma$ uncertainties of the $\chi^2$-fitting. Since the cadence after 115 days is poor, these observations were plotted separately.}
    \label{fig:bolfit}
\end{figure}

We constructed a pseudo-bolometric LC of SN~2021efd by fitting its optical spectral energy distributions (SEDs) with single blackbody (BB) functions with both radius and temperature set as free parameters. The fitting was done using the linearly interpolated $uBVcgroi$ data. We only used the epochs that had detections in three or more bands within 10 days (5 days before or after them). We determined the best fits using the $\chi^2$ goodness-of-fit test, and the errors represent the 1-$\sigma$ uncertainties. The results are displayed in Fig.~\ref{fig:bolfit}.
We note that, since the radiation at late phases does not exhibit a BB spectrum but is instead dominated by emission lines (see also Sect.~\ref{sec:spec_properties}), the inferred BB parameters at late phases should be considered reference values based on the assumption of BB radiation.

The pseudo-bolometric LC to first order follows the evolution of the LC in the single bands, also displaying multiple bumps and an overall slow decline. The peak luminosity during the first peak is $\sim 2.8\times10^{42}~\mathrm{erg}$. The total radiated energy derived from the pseudo-bolometric LC until $\sim 710$ days is $\sim 4.9\times 10^{49}~\mathrm{erg}$. The total generated energy from the decay of $^{56}$Ni and $^{56}$Co is $1.89\times 10^{50}\times M_{\mathrm{Ni}}/M_\odot~\mathrm{erg}$ \citep[][]{arnett1980,arnett1982} and $M_\mathrm{Ni}$ for SESNe ranges from $\sim 0.01$ to $0.3 ~M_\odot$ \citep[e.g.,][]{taddia18,afsariardchi21}, thus the typical energy released in radioactive decay is $\sim 10^{48}-10^{49}$ erg. Due to incomplete $\gamma$-ray trapping, not all of the released energy contributes to the observed luminosity.

The BB temperature evolution shows cooling until phase 25 days. The sudden drop in temperature happened at the time of the last observations with the Swope telescope (27 days), which is probably caused by the end of the coverage on the $uBV$ bands. After this, the temperature appears to be roughly constant. Later, at $\gtrsim$ 300 days from the $r$-band peak, SN~2021efd shows an increase in temperature, although at later epochs the error on the temperature is large. During the photospheric phase, the SN is well described by a BB function, but as the SN ejecta expands, the SN becomes emission-line dominated. At later epochs, this increases the BB fit uncertainties, especially in the calculated temperature, which is sensitive to the relative brightness in different bands. In addition to this, the cadence of the observations drops after $\sim 115$ days. The BB fit at later phases $\gtrsim100$ days may not reflect the actual physical characteristics of the SN, and the derived values are indicative.

The BB radius increases to $\sim 2 \times 10^{15}$~cm toward the LC peak and remains roughly constant for $\sim 70$~days. It slightly decreases to $\sim 1.5 \times 10^{15}$~cm as the BB temperature increases at $\sim 80$ days after the $r$-band peak. After the gap in observations, the radius is relatively stable at $\sim 2 \times 10^{14}$~cm. From the initial rise of the BB-radius we can roughly estimate a photospheric velocity of $\sim6000$ km s$^{-1}$ measured from -12 days until -5 days.

\section{Spectroscopic properties} \label{sec:spec_properties}

\subsection{Spectroscopic evolution}

\begin{figure}
        \begin{center}
        \includegraphics[width=0.90\linewidth]{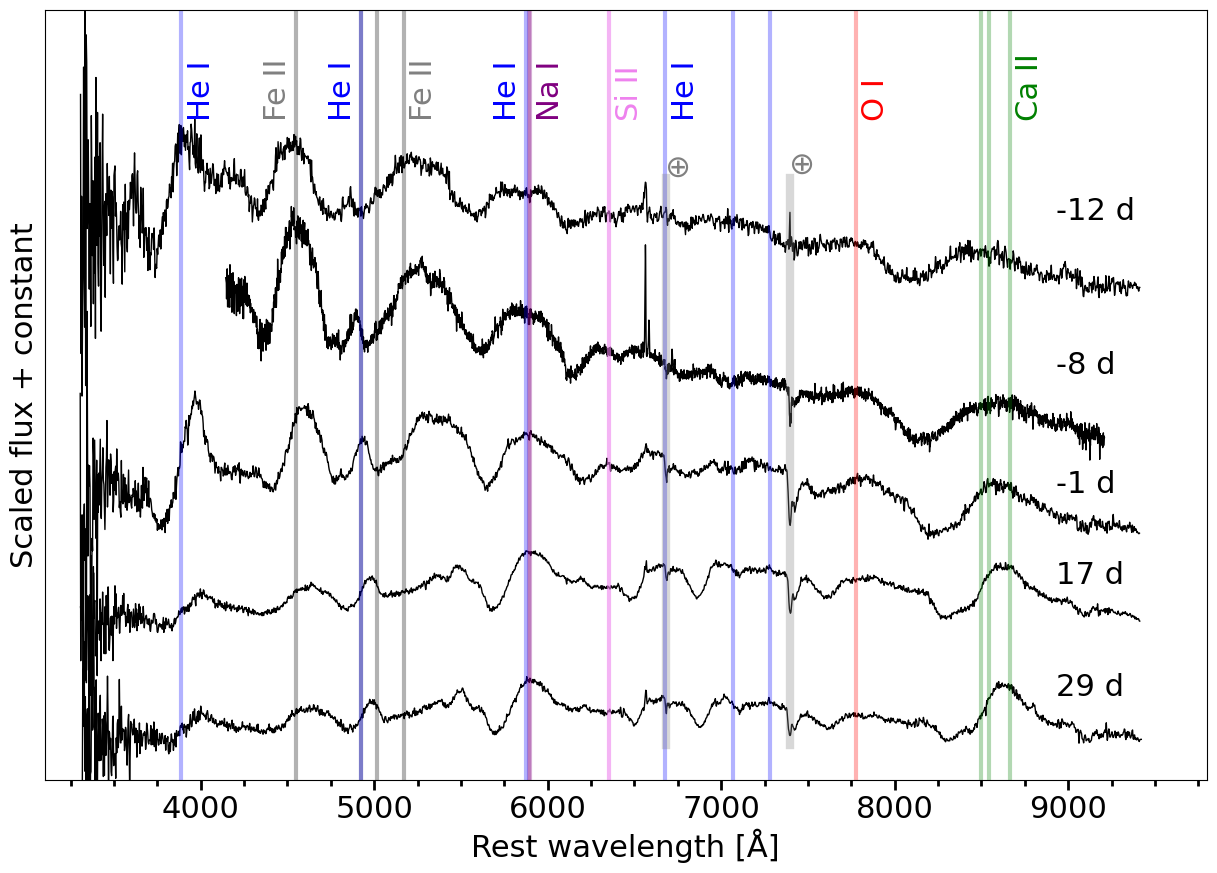}
        \caption{Early-time spectra of SN~2021efd. The flux scale is arbitrary. The telluric features are marked with grey lines. The locations of some typical lines at zero velocities are indicated.}
        \label{spec}
        \end{center}
\end{figure}

\begin{figure*} [ht]
    \centering
    \includegraphics[width=0.90\linewidth]{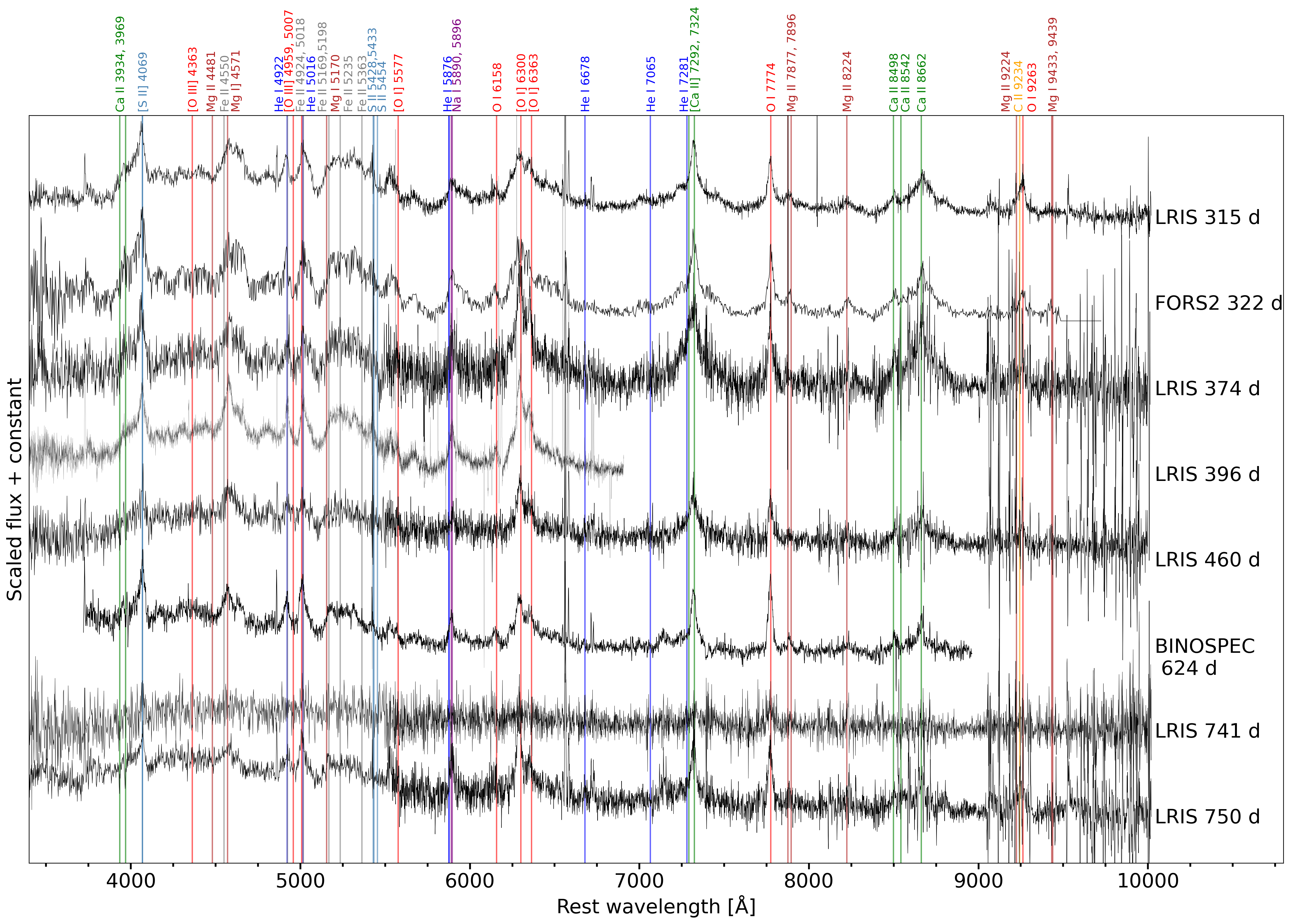}
    \caption{The nebular phase spectra of SN~2021efd. The spectra are corrected for telluric features. Locations of some typical lines are indicated in the Figure, even if they are not present.}
    \label{nebular_spectra}
\end{figure*}

\begin{figure} 
    \centering
    \includegraphics[width=0.99\hsize]{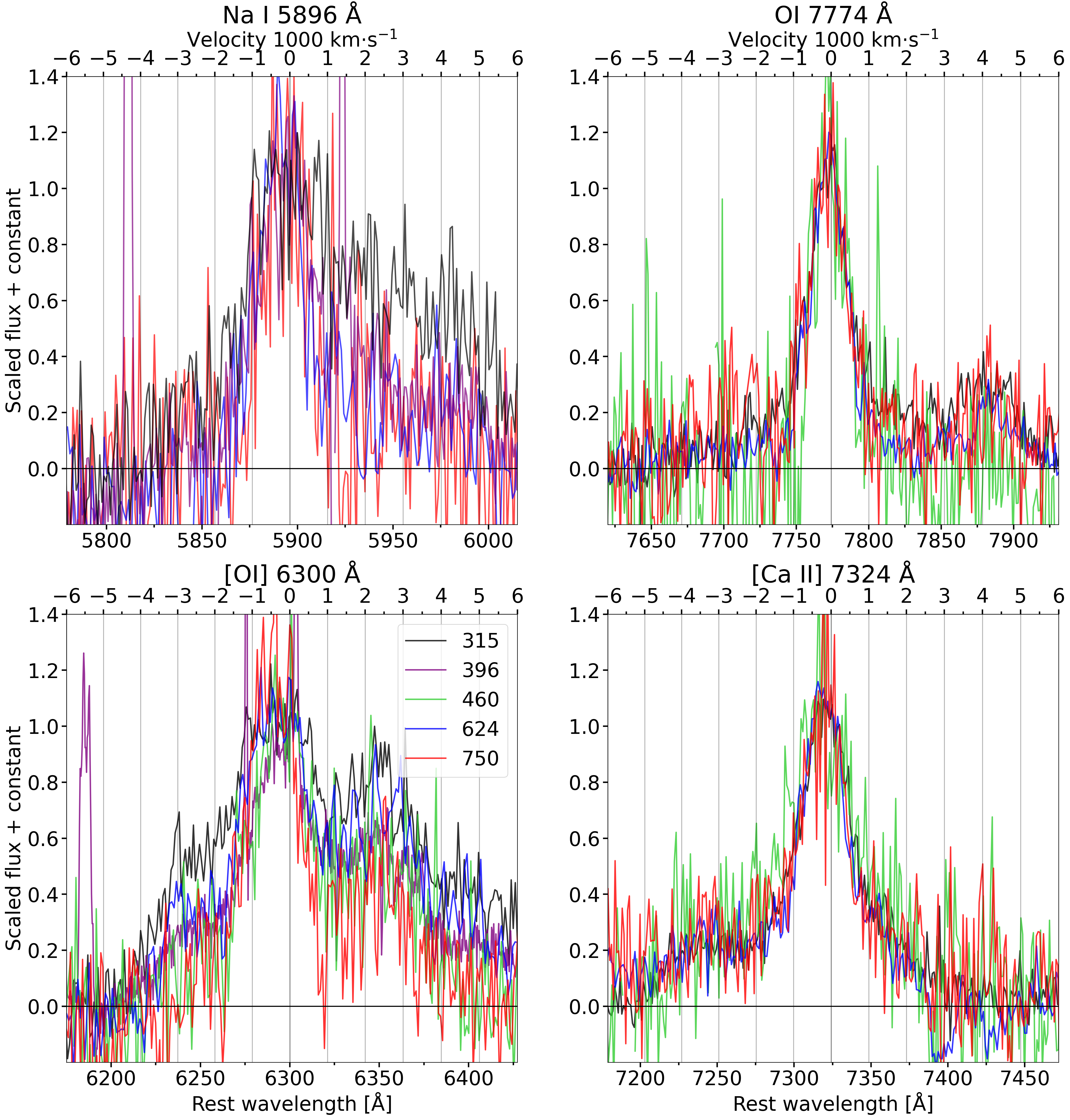}
    \caption{The line profiles of the nebular spectra. Some epochs are excluded for clarity. The rest wavelength of what appears to be the strongest line is displayed as the zero point of velocity.}
    \label{fig_lineprof}
\end{figure}

The early-phase spectra of SN~2021efd are displayed in Fig.~\ref{spec}. They were taken before the rebrightening, which begins at around 60 days. These spectra are typical of the early and near-maximum phases of Type Ib SNe with \ion{He}{i}, \ion{Fe}{ii}, and \ion{Ca}{ii} lines being present. A weak \ion{Si}{ii} is also present, and a \ion{O}{i} line starts forming at 7774~\AA\ after the brightness peak. The narrow H$\alpha$-[\ion{N}{ii}] lines in the spectra likely originate from the host galaxy.

The nebular spectra of SN~2021efd, displayed in Fig. \ref{nebular_spectra}, are dominated by emission lines, as is typical for SESNe. The [\ion{O}{i}] doublet at $\sim$6300 \AA, the \ion{O}{i} line at 7774~\AA, the  [\ion{Ca}{ii}] doublet at $\sim 7200$~\AA, and the  \ion{Ca}{ii} NIR triplet at $\sim8500$~\AA\ are the most prominent features. Additionally, He lines are present at 4921~\AA\ and 5876 \AA. These coincide in wavelength with an \ion{Fe}{ii} and a \ion{Na}{i} line, respectively, but as there are other prominent \ion{He}{i} lines present, we consider the \ion{He}{i} identification as more likely. The blue part is dominated by \ion{Fe}{ii} complex lines. At 4571~\AA\ there is also a superposition of \ion{Mg}{i}] and [\ion{Fe}{ii}]. In the nebular spectrum, prominent lines (e.g., \ion{O}{i}, \ion{Ca}{ii}) have both broad and narrow ($\sim$1000 km~s$^{-1}$) components, which is observed in other interacting SESNe as well (e.g., SN~2022xxf, \citealt{kuncarayakti23}). The time evolution of the lines in the nebular spectra is displayed in Fig. \ref{fig_lineprof}. Except for some small variation, the nebular lines remain mostly unchanged.

\subsection{Spectral comparison}

\begin{figure}
    \centering
    \includegraphics[width=\hsize]{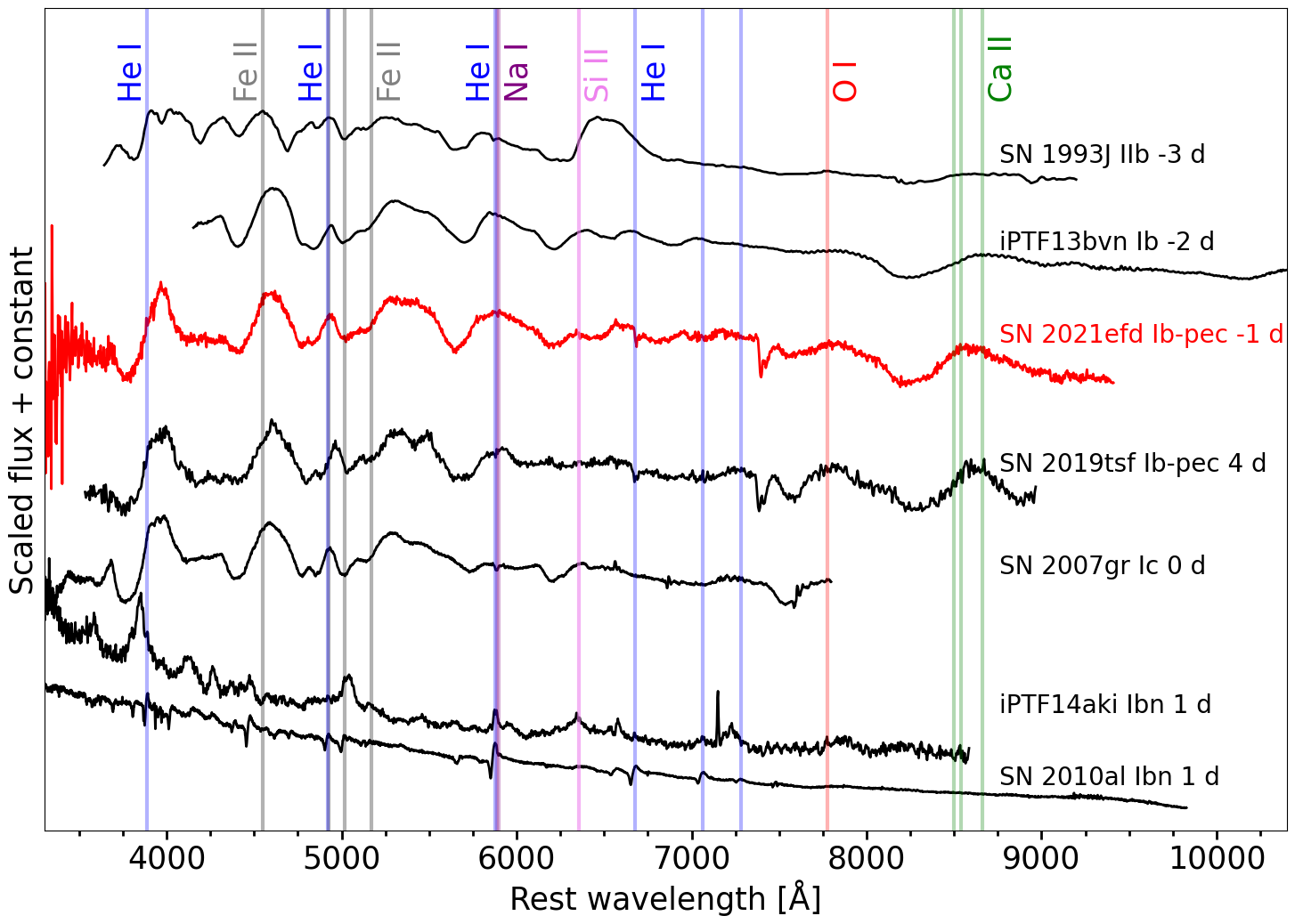}
    \caption{The photospheric phase spectra of SN~2021efd and multiple other SESNe. Flux normalization was done by dividing the flux at every wavelength by the peak flux between $4200-5000$~\AA. The locations of some typical lines are indicated.}
    \label{early comp}
\end{figure}

The early-phase spectra of SN~2021efd are compared in Fig.~\ref{early comp} with those of the Type~Ib iPTF13bvn \citep{cao13}, the Type~IIb SN~1993J \citep{barbon95},  the Type~Ic SN~2007gr \citep{valenti08}, the peculiar Type~Ib SN 2019tsf \citep{sollerman20}, as well as with  the Type~Ibn SNe iPTF14aki \citep{hosseinzadeh17} and 2010al \citep{pastorello15}.
At early phases, SN~2021efd is similar to the typical Type Ib SNe. All the lines present in SN~2021efd are also present in iPTF13bvn. The other SESNe differ notably by their lack of helium in the case of SN~2007gr, or by the presence of hydrogen in SN~1993J. SN~1993J also has weaker \ion{Fe}{ii} lines in the blue part of the spectrum and a weaker \ion{Ca}{ii} NIR triplet absorption feature. Observationally SN~2021efd has no resemblance to Type Ibn SNe. The spectra of Type~Ibn SNe 
are mostly continuum-dominated, with iPTF14aki showing a greater number of narrow lines in its spectra.
The spectral features of SN~2019tsf are similar to those of SN~2021efd and iPTF13bvn, which is noteworthy as its LC decline rate at this epoch already suggests a contribution from another powering mechanism in addition to radioactive decay.

\begin{figure} 
    \centering
\includegraphics[width=\hsize]{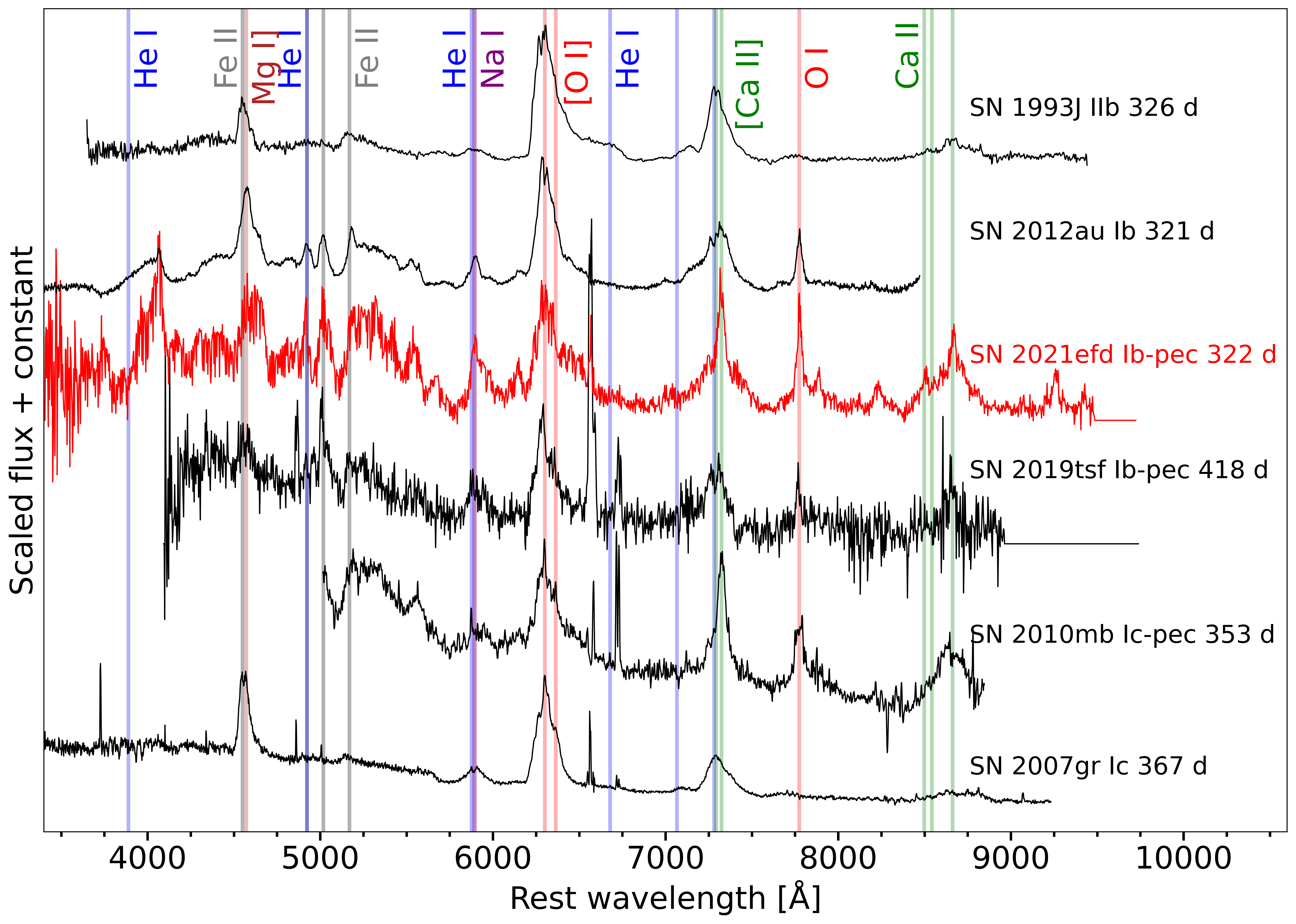}
    \caption{The nebular phase spectra of SN~2021efd and several other SESNe. Flux normalization was done by dividing the flux at every wavelength with the peak flux of the [\ion{O}{i}] line at $\sim 6300$~\AA.}
    \label{neb comp}
\end{figure}

The nebular-phase spectrum of SN~2021efd is compared in Fig.~\ref{neb comp} to similar phase spectra of the comparison sample. These include  
SN~1993J \citep{jerkstrand14}, the Type~Ib SN~2012au \citep{milisavljevic13}, SN~2007gr \citep{shivvers}, SN~2019tsf and the interacting Type~Ic SN~2010mb \citep{ben-ami14}.
The spectrum of SN~2021efd exhibits a narrow \ion{O}{i} $\lambda7774$ line similar to the one observed in SN~2019tsf. It can also be seen that the \ion{Ca}{ii} NIR triplet is much stronger in SN~2021efd than in the regular SESNe. Similar to SN~2019tsf, SN~2021efd has an excess of continuum flux around the iron complex ($\sim 4000 - 5500$~\AA). This may suggest a similar powering mechanism of H-free CSI. Despite the lack of narrow lines in the spectrum of SN~2019tsf, they can not be ruled out due to the noisiness of the spectra. 
SN~2010mb also exhibits a blue excess and strong \ion{Ca}{ii} NIR lines.

\subsection{Velocity evolution}

\begin{figure} 
    \centering
\includegraphics[width=\hsize]{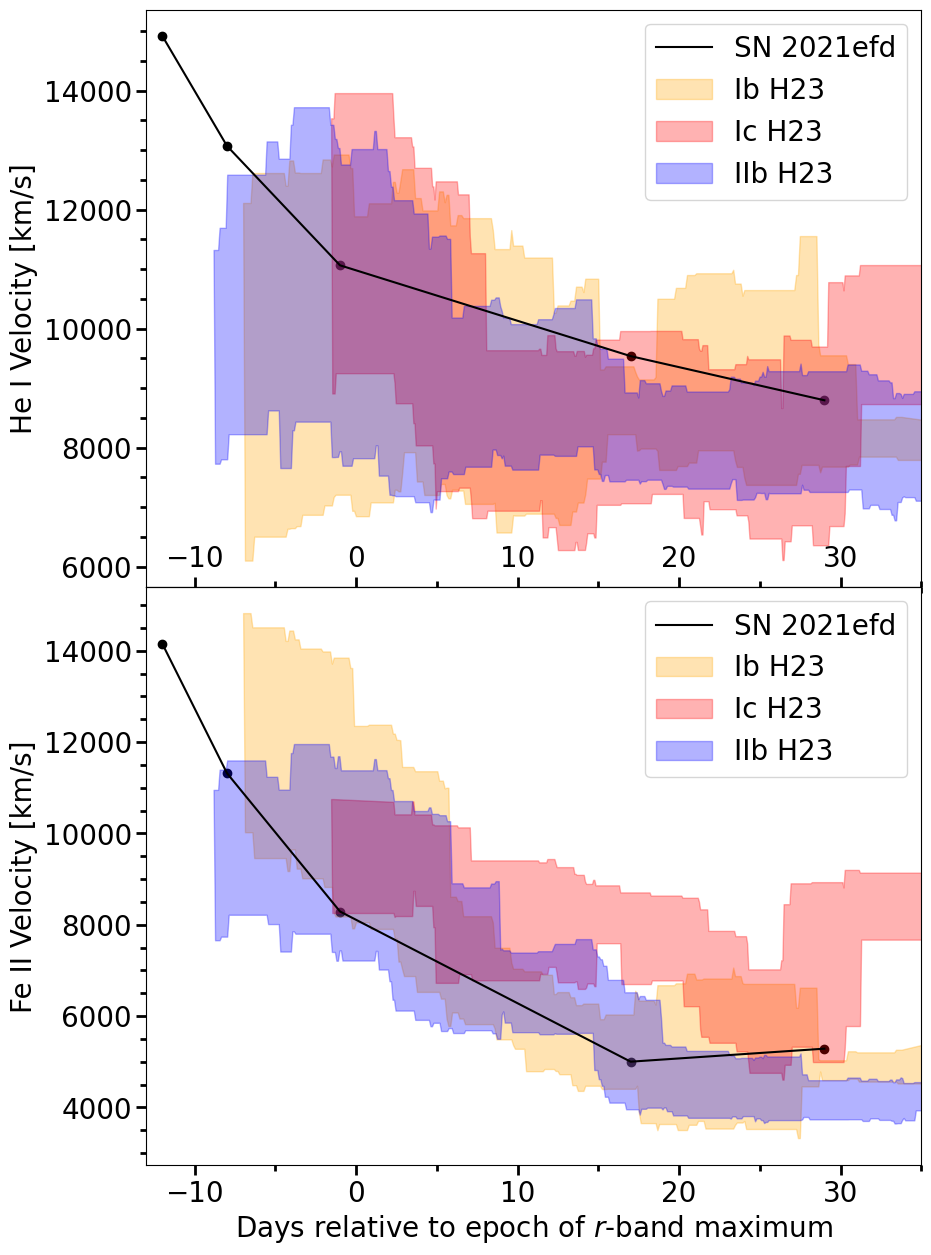}
    \caption{\textit{Top:} \ion{He}{i} $\lambda5876$ velocity of SN~2021efd. \textit{Bottom:} \ion{Fe}{ii} $\lambda5169$ velocity of SN~2021efd. The templates are the rolling mean of SESNe as presented by \citet[][]{holmbo23}; H23. The $\lambda$5876  feature they measured for Type~Ic SNe is expected to be composed of \ion{Na}{i} $\lambda\lambda$5890,5896 doublet, and little to no \ion{He}{i} $\lambda5876$.}
    \label{vel}
\end{figure}

We calculate the velocity of the \ion{Fe}{ii} $\lambda 5169$ line and the \ion{He}{i} $\lambda5876$ line from their absorption minima by fitting the absorption features with a Gaussian function. 
The \ion{He}{i} 5876~\AA\ feature might have a contribution from \ion{Na}{i} $\lambda\lambda5890, 5896$.
In Fig.~\ref{vel}, we compare the spectral line evolution of these features along with those measured from the  SESN sample published by the Carnegie Supernova Project \citep[][]{stritzinger2023,holmbo23}. 
The ejecta velocity of SN~2021efd inferred from both the \ion{He}{i} and \ion{Fe}{ii} lines in the spectra at the early epochs is in the range of what is typical for SESNe (Fig.~\ref{vel}). While the \ion{He}{i} velocity is around the higher end for an  SESN, the \ion{Fe}{ii} velocity is at the lower end, though both velocities are still within the distributions.
The velocity inferred from the early-time evolution of the BB radius ($\sim 6000$ km~s$^{-1}$; see Sect.~\ref{sec:phot_properties}) is roughly in line with the \ion{Fe}{ii} velocity measured from the spectra ($\sim 8000$ km~s$^{-1}$ around the brightness peak).

\subsection{[\ion{O}{i}]/[\ion{Ca}{ii}] line ratio} \label{sec:OI/Ca}
 
We measured the fluxes of the [\ion{O}{i}] $\lambda\lambda6300,6364$ and the [\ion{Ca}{ii}] $\lambda\lambda$7292, 7324 doublets. We determine a continuum level and then subtract the nearby \ion{Fe}{ii} and \ion{N}{ii} lines by assuming a symmetric profile, as done in \citealt{fang22}.
The [\ion{O}{i}] to [\ion{Ca}{ii}] ratio of SN~2021efd is 1.3-1.4 during the epochs from 324 days to 473 days. At 641 days and at 771 days the ratio is 1.7. 
The measurement of the [\ion{O}{i}] to [\ion{Ca}{ii}] ratio might be affected by uncertainties from the nearby lines, in particular their possible asymmetry due to the assumption of the symmetric profiles of \ion{Fe}{ii} and \ion{N}{ii}. 
It has been found that Type~IIb and Type~Ib SNe have similar [\ion{O}{i}]/[\ion{Ca}{ii}] flux ratios, while Type~Ic SNe typically have higher values, although overlapping with SNe~IIb/Ib \citep{fang22}. 
Compared to the values reported in Fig. 3 of \citet{fang19}, SN~2021efd sits in the overlapping region of SESNe.

\section{Supernova ejecta properties} \label{sec:SN_properties}

We estimate the ejecta properties of SN~2021efd, such as the mass, kinetic energy, and $^{56}$Ni mass of the ejecta, under the assumption that the CSM interaction has a minimal contribution to the SN brightness before and around the first LC peak. This assumption is supported by the fact that the early-phase photometric and spectroscopic properties of SN~2021efd are similar to those of the other, non-interacting, SESNe.

For the estimation of the mass and kinetic energy of the ejecta, we use the following equations in the one-zone analytic model for the LCs from the homologously expanding gas with radiative diffusion \citep{arnett1982, arnett1996}:
\begin{align*}
M_{\text {ej }}=\frac{1}{2} \frac{\beta \mathrm{c}}{\kappa} v_{p h} t_{\max }^2,
\end{align*}
\begin{align*}
E_k=\frac{3}{20} \frac{\beta \mathrm{c}}{\kappa} v_{ph}^3 t_{\max }^2,
\end{align*}
where $\beta$ is an integration constant equal to 13.8, $\kappa$ is the opacity of the ejecta gas, $v_{ph}$ is the ejecta velocity, and $t_{\max}$ is the rise time of the LC. Here, $\kappa$ is assumed to be 0.1 $\mathrm{cm}^2~\mathrm{g}^{-1}$, which is typical for ionized H-poor gas, although it can vary depending on the chemical composition of the gas. We assume the ejecta velocity to be $8300$ km s$^{-1}$, which is the velocity inferred from the \ion{Fe}{ii} $\lambda$5169 line around the light curve peak, and the rise time to be 18.5~days, which is the value estimated from the pseudo-bolometric LC.
We find the ejecta mass to be  $2.2 \pm0.3~M_\odot$ and the kinetic energy of the ejecta to be $9.1 \pm1.2\times 10^{50}$ erg, where the uncertainty is due to the error of the rise time. 

We estimate the $^{56}$Ni mass from its relation with the peak luminosity \citep{arnett1980, arnett1982}:
\begin{small}
\begin{equation*}
    L (t_{\rm{max}}) =\left(6.45 \times 10^{43} e^{-t_{\rm{max}} / 8.8~\mathrm{d}}+1.45 \times 10^{43} e^{-t_{\rm{max}} / 111.3~\mathrm{d}}\right) M_{\mathrm{Ni}}~ \mathrm{erg}~ \mathrm{s}^{-1}.
\end{equation*}  
\end{small}
This gives an estimate of $0.14 \pm 0.01 ~M_\odot$, where the error is calculated using the limits from the bolometric luminosity fit. This method is known to overestimate the $^{56}$Ni mass by a factor of $\sim 2$ \citep{afsariardchi21}. 
In addition, since the peak luminosity of SN~2021efd might be enhanced by the contribution of potential interaction with CSM, the actual $^{56}$Ni mass can be smaller than this estimated value.
The total energy released by the radioactive decay of this amount of $^{56}$Ni is $(2.6 \pm 0.2)\times 10^{49}~\mathrm{erg}$, which is 
smaller than the total energy radiated by SN 2021efd ($4.9 \times 10^{49}~\mathrm{erg}$) until 710 days, as measured by integrating the entire pseudo-bolometric LC.

We compare these derived values of the ejecta mass, kinetic energy, and $^{56}$Ni mass for SN 2021efd to those of other SESNe. \citet{taddia18}, who followed a similar approach in their work, found the average values for Type Ib SNe to be $M_\mathrm{ej}=3.9 \pm2.7~M_\odot$, $E_\mathrm{k}=1.6 \pm 1.0 \times 10^{51}$ erg, and $M_\mathrm{Ni}=0.10 \pm 0.05~M_\odot$ respectively. For Type Ic, they estimated $M_\mathrm{ej}=2.5 \pm1.2~M_\odot$, $E_\mathrm{k}=1.2 \pm 0.8 \times 10^{51}$ erg, and $M_\mathrm{Ni}=0.14 \pm 0.04~M_\odot$. The errors here are the standard deviations of the sample. The values that we derived for SN 2021efd ($M_\mathrm{ej}=2.2 \pm0.3~M_\odot$, $E_\mathrm{k}=0.91 \pm 0.12 \times 10^{51}$ erg, and $M_\mathrm{Ni}=0.14 \pm 0.01~M_\odot$) are consistent with these average values for Type~Ib and Type~Ic SNe within the standard deviation. While SN~2021efd has slightly lower ejecta mass and kinetic energy than the average values, its $^{56}$Ni mass is slightly higher than the average value for Type~Ib SNe and similar to the average value for Type~Ic SNe.

\section{Origin of the late-time excess and the CSM properties} \label{sec:origin}

In this section, we discuss the origin of the late-time excess in the brightness of SN~2021efd and derive its CSM properties with LC modeling. 
Several possible scenarios for the late-phase rebrightening in SESNe \citep[e.g.,][]{moore23, kuncarayakti23, chen24, kangas24} have been proposed. These include: delayed CSM interaction \citep[e.g.,][]{chevalier06}, delayed activity of a newly formed magnetar \citep[e.g.,][]{kasen10}, or delayed accretion onto a newly formed compact object \citep[e.g.,][]{dexter13}.
In the case of SN~2021efd, the LC shows clear bumps at $\sim75$ and $\sim100$ days from the explosion, and possibly at 325, 345, 390, and 400 days as well. With either the magnetar or fall-back accretion scenarios, it is difficult to explain not only the delayed onsets of the rebrightenings but also the multiple onsets \citep[e.g.,][]{kasen10, dexter13}. The multiple bumps in the LC favor the CSM interaction scenario, where the bumps are created by interactions with multiple shell-like or clumpy CSM.
At the same time, since its early-time photometric and spectroscopic properties are similar to those of prototypical Type~Ib SNe (see Sect.~\ref{sec:phot_properties} and \ref{sec:spec_properties}), it might be natural to consider the cause of the rebrightenings as not the difference of the SN ejecta or progenitors (e.g., as in the magnetar or fall-back accretion scenarios) but the difference of some external factor (e.g., as in the CSM interaction scenario).
Furthermore, the features in the late-time spectrum (i.e., the spectra from phase 324 days onward; see Fig.~\ref{nebular_spectra}) support the CSM interaction scenario. 
The narrow emission lines with velocities of $\sim$1000 kms$^{-1}$ point to slower-moving material than the bulk SN ejecta, consistent with the CSM interaction scenario. A similar velocity scale was observed in post-maximum near-infared spectra of  Type~Ic SN~2016adj which also showed signatures of interaction with H-rich gas \citep{Stritzinger2024}.
The so-called Fe bump near $\sim$5200 \AA, likely created by highly ionized and/or excited Fe lines, is frequently seen in interacting SNe of various types, including Type~IIn, Ia-CSM, interacting Ibc, Ibn, and Icn \citep[e.g.,][]{sharma23,kuncarayakti18,pastorello08,pellegrino22}.

Therefore, we conclude that the late-time LC bumps in SN~2021efd are due to interaction with multiple-shell, multiple-ring, or clumpy CSM.

\subsection{Light curve model}

In order to derive the properties of the CSM in SN~2021efd, we analyzed the bolometric LC using the luminosity model presented in \citet{moriya13}.
The strong interaction between the SN ejecta and CSM creates a shock shell, in which the kinetic energy of some parts of the ejecta is transformed into radiation energy. For calculating the radiation from the CSM interaction, we assess the evolution of the shock shell. We assume that the parameters within the shock shell are uniform, and the width of the shock shell is negligible in comparison to its extent.
As in \citet{moriya13}, we derive the time evolution of the location ($r_{\mathrm{sh}}$) and velocity ($v_{\mathrm{sh}}$) of the interaction shock shell using its equation of motion:
\begin{align}
M_{\mathrm{sh}} \frac{\mathrm{d} v_{\mathrm{sh}}}{\mathrm{d} t}=4 \pi r_{\mathrm{sh}}^2\left[\rho_{\mathrm{ej}}\left(v_{\mathrm{ej}}-v_{\mathrm{sh}}\right)^2-\rho_{\mathrm{csm}}\left(v_{\mathrm{sh}}-v_{\mathrm{csm}}\right)^2\right],
\end{align}
where $\rho$ and $v$ are density and velocity, respectively, and the subscripts of ej, sh, and csm correspond to the ejecta, shock shell, and CSM, respectively. $M_{sh}$ is the total mass of the shock shell, i.e., the shocked CSM and ejecta. We assume that the CSM velocity is constant, and that the ejecta is homologously expanding: $v_\mathrm{ej}(r)=${\Large $\frac{r}{t}$}. We use the density profiles of the ejecta that are estimated based on numerical simulations \citep{matzner1999}, i.e., the double power-law distribution of \cite{moriya13}:
\begin{align}
    \rho_{\mathrm{ej}}\left(v_{\mathrm{ej}}, t\right)= \begin{cases}\frac{1}{4 \pi(n-\delta)} \frac{\left[2(5-\delta)(n-5) E_{\mathrm{ej}}\right]^{(n-3) / 2}}{\left[(3-\delta)(n-3) M_{\mathrm{ej}}\right]^{(n-5) / 2}} t^{-3} v_{\mathrm{ej}}^{-n} & \left(v_{\mathrm{ej}}>v_t\right), \\ \frac{1}{4 \pi(n-\delta)} \frac{\left[2(5-\delta)(n-5) E_{\mathrm{ej}}\right]^{(\delta-3) / 2}}{\left[(3-\delta)(n-3) M_{\mathrm{ej}}\right]^{[\delta-5) / 2}} t^{-3} v_{\mathrm{ej}}^{-\delta} & \left(v_{\mathrm{ej}}<v_t\right),\end{cases}
\end{align}
where $\delta$ and $n$ are the outer and inner density slopes, respectively. $M_{\mathrm{ej}}$ and $E_{\mathrm{ej}}$ are the mass and energy of the ejecta, respectively, and $v_t$ is the velocity at the border of the power laws: 
\begin{align}
v_t=\left[\frac{2(5-\delta)(n-5) E_{\mathrm{cj}}}{(3-\delta)(n-3) M_{\mathrm{ej}}}\right]^{\frac{1}{2}}.
\end{align}

For the mass and energy of the ejecta, we adopt the values estimated in Sect. \ref{sec:SN_properties}: $M_{\mathrm{ej}}=2.2 \pm 0.3$ M$_\odot$ and $E_{\mathrm{ej}}=9.1 \pm 1.2 \times 10^{50}$ erg.
We adopt the values $n=10$ and $\delta=1$ \citep[e.g.,][]{matzner1999,kasen10b,moriya13}.

We assume the rate of the mass loss from the progenitor system ($\dot{M}$) is constant in time, and thus the density of the CSM at distance $r$ is written as:
\begin{align}\label{eq3}
    \rho_\mathrm{CSM}(r)&=\frac{\dot{M}}{4\pi v_\mathrm{csm} r^2} = D r^{-2}.
\end{align}

Here, we define $D$ as $\dot{M}/(4\pi v_{\rm{csm}})$. Since we do not know the mechanism of the mass loss, the expansion velocity of the CSM ($v_\mathrm{csm}$) is unclear. We estimate the CSM velocity to be $\sim100 -1000~\mathrm{km}~\mathrm{s}^{-1}$ (see Sect.~\ref{sec:mass-loss_properties}). Here, we simply use the minimum value of $v_\mathrm{csm}=100~\mathrm{km}~\mathrm{s}^{-1}$, which produces the minimum value of the necessary mass-loss rate, as a fiducial value, and also discuss the case with the maximum value of $v_\mathrm{csm}=1000~\mathrm{km}~\mathrm{s}^{-1}$.

By numerically solving the equation of motion with the ejecta and CSM parameters assumed above, we calculate the distance and velocity of the shock shell as a function of time. 
We assume the optically thin limit, where a fraction of the generated energy ($\frac{\mathrm{d} E_{\mathrm{kin}}}{\mathrm{d} t}$) immediately escapes from the shocked shell as optical radiation, and calculate the optical luminosity as follows \citep{moriya13}:
\begin{equation}
L=\epsilon \frac{\mathrm{d} E_{\mathrm{kin}}}{\mathrm{d} t}=2 \pi \epsilon \rho_{\mathrm{csm}} r_{\mathrm{sh}}^2 v_{\mathrm{sh}}^3.
\end{equation}
The value $\epsilon$ is the conversion efficiency from the kinetic energy of the SN ejecta into the optical radiation due to the CSM interaction. We assume the value of $\epsilon$ to be 0.1 \citep{moriya13b}.

\subsection{Comparison with the observation}

\begin{figure}
    \centering
    \includegraphics[width=\hsize]{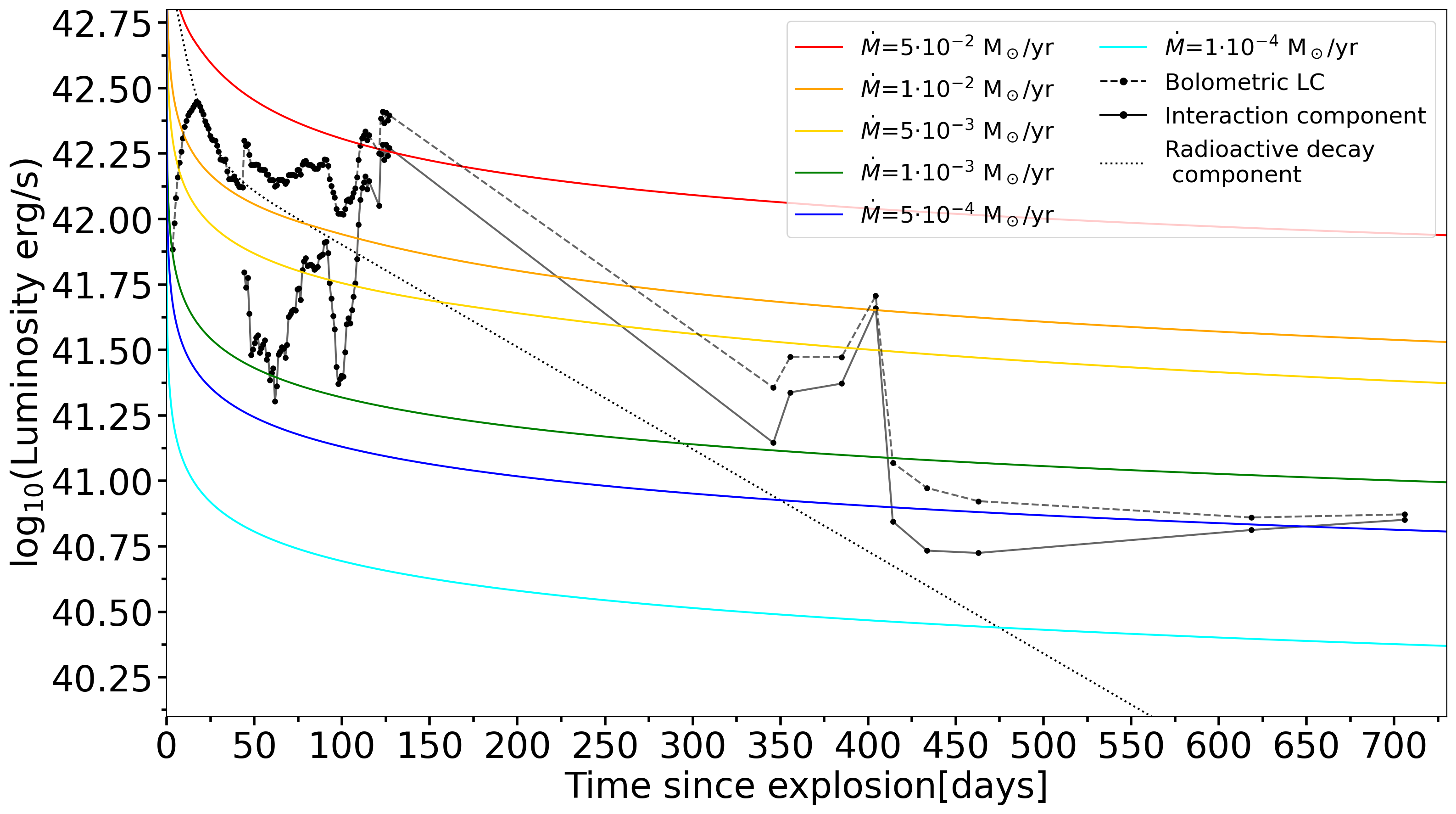}
    \caption{Comparison of SN~2021efd bolometric light curve to the models.
    The colored lines are the calculated luminosities from interaction based on the model. The black line is the bolometric luminosity of SN~2021efd with the radioactive decay component subtracted. The dashed line is the observed bolometric LC, and the dotted line is the radioactive decay component calculated from peak luminosity.}
    \label{model_comp}
\end{figure}

The bolometric LC of SN~2021efd is compared to the calculated interaction luminosity in Fig. \ref{model_comp}. The radioactive-decay component is subtracted using the analytical formula presented in \citet{arnett1980, arnett1982} and assuming the $^{56}$Ni mass to be 0.14 M$_\odot$. Since this formula assumes full $\gamma$-ray trapping, the later parts of this radioactive-decay LC are likely overestimated. The actual interaction component is thus expected to be somewhere between the LC without the radioactive component and the original bolometric LC.
After $\sim 60$ days from the explosion onward, the interaction luminosity is increasing. This could be due to an increase in the corresponding mass-loss rate for the encountered CSM, suggesting the presence of an inner cavity in the CSM distribution. The early parts (at least from $\sim 80$ to $\sim 120$ days) can be roughly explained by the LC models with $\sim 5\times 10^{-3}- \sim 10^{-2}~M_\odot~\mathrm{yr}^{-1}$. The later parts (at least after $\sim 340$ days) are significantly lower than these levels, which suggests that the mass-loss rate of the progenitor of SN~2021efd is not constant, but grows with time toward the explosion.

Here, we note a caveat on the estimated mass-loss rate. The CSM velocity is directly proportional to the mass-loss rate, as can be seen from Eq. \ref{eq3}. CSM velocities of 1000-2000 km s$^{-1}$ have been found for Type Ibn SNe, and have been suggested for WR stars \citep[e.g.,][]{hosseinzadeh17, crowther07}. A velocity of 1000 km s$^{-1}$ would increase the estimated mass-loss rate by an order of magnitude, to $10^{-1}~M_\odot~\mathrm{yr}^{-1}$. Since we cannot estimate the precise CSM velocity in the case of SN~2021efd, the estimated mass-loss rate is a lower limit. The uncertainty of the value of the efficiency $\epsilon$, which is directly proportional to the luminosity, does not largely affect the estimated value. In fact, even a change of the assumed $\epsilon$ value by a factor of two, i.e., $\epsilon=0.05-0.2$, does not change the order of magnitude of the estimate, as this is equivalent to adding or subtracting $\mathrm{log}_{10}(L)$ by $\mathrm{log}_{10}(2)=0.3$ (see Fig. \ref{model_comp}).

\begin{figure}
    \centering
    \includegraphics[width=0.5\textwidth]{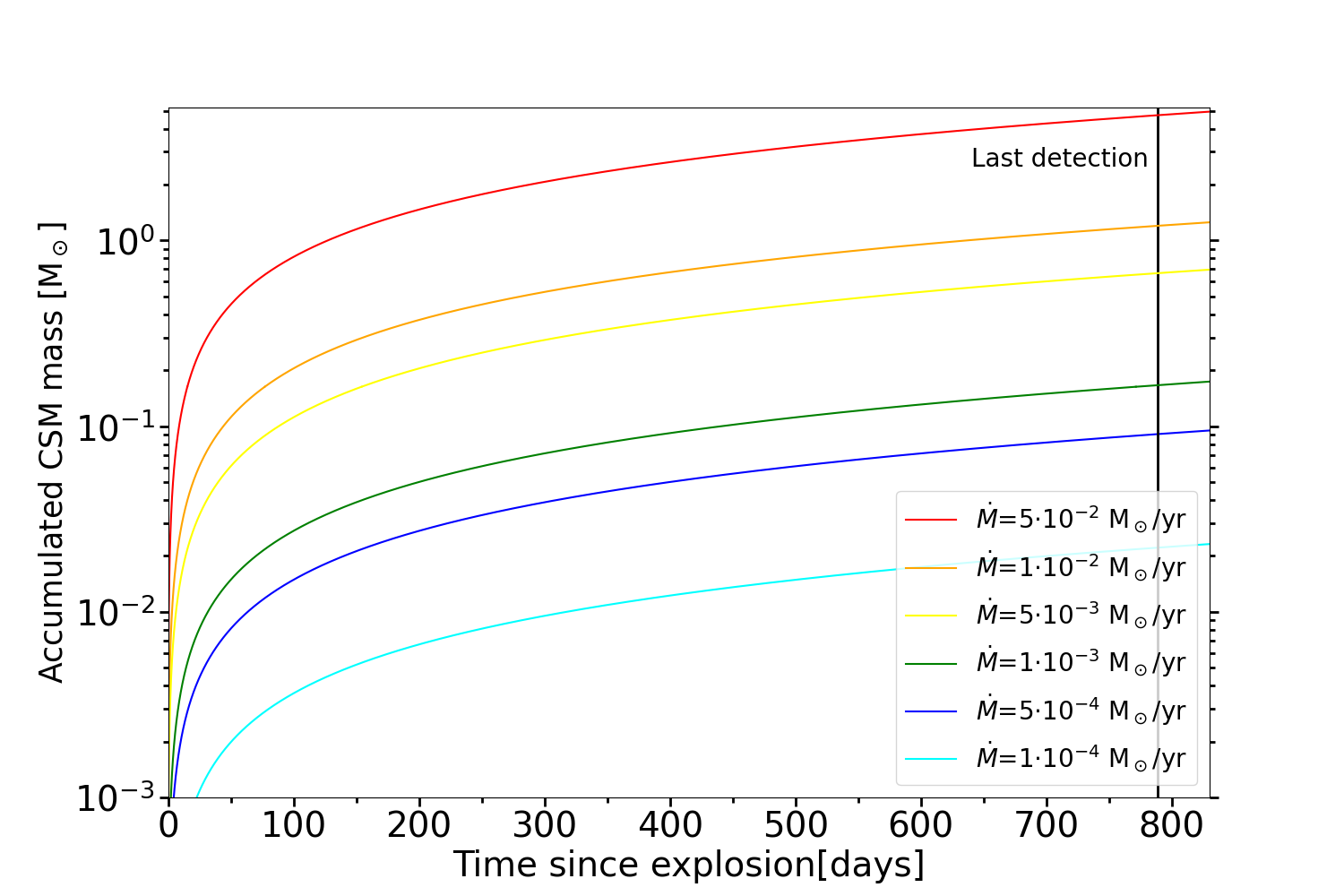}
    \caption{The mass of the shock shell accumulated from the CSM at each epoch according to the values used in the calculated interaction luminosity. Effectively, this tells us the CSM mass up to a certain distance from the progenitor star.}
    \label{accmass}
\end{figure}

Figure \ref{accmass} displays the accumulated mass of the shock shell over time. From this figure, we can see that the masses for the good-fit models already reach a few times $10^{-1}~M_\odot$ at $\sim 120$ days. This is the lowest limit of the actual total CSM mass, as in the observed LCs we can see excessive flux until at least $\sim 770$ days and possibly later as well (see Fig. \ref{lc}).

\subsection{CSM geometry} \label{sec_CSM_geometry}

\begin{figure} [t]
    \centering
    \includegraphics[width=0.5\textwidth]{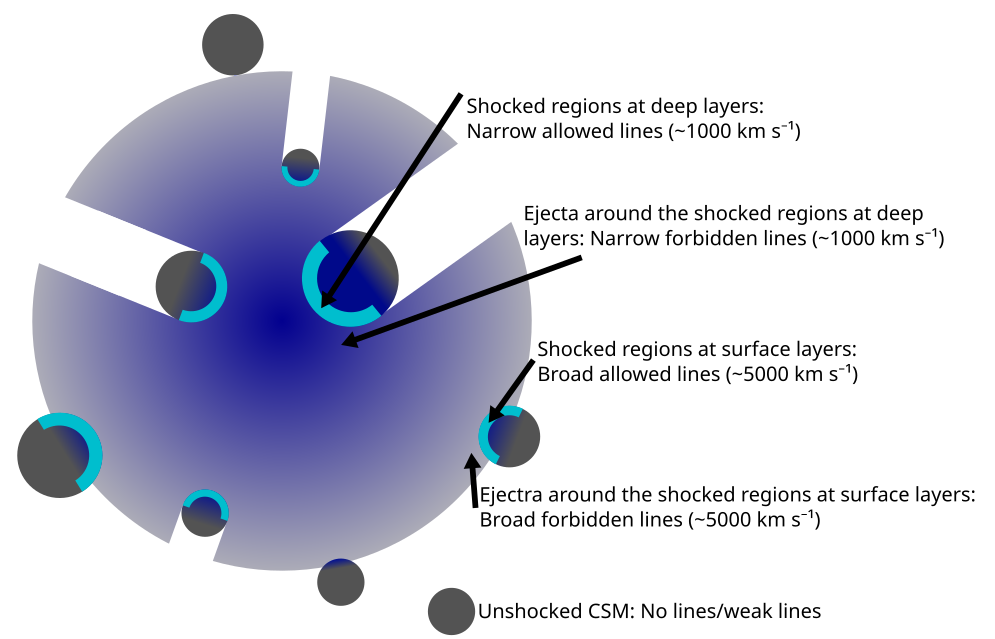}
    \caption{Schematic of the preferred model in the nebular phase. Arrows are used to mark characteristic regions for the narrow and broad components, respectively. The light blue coloring approximately corresponds to the shock regions.}
    \label{schematic}
\end{figure}

Here we discuss the geometry of the CSM, which is directly related to the mass-loss mechanism. We have concluded that the late-time rebrightenings in SN~2021efd are due to interaction with CSM with a multiple-shell/ring-like or clumpy structure. The LC modeling demonstrates that mass loss with an average rate of $\dot{M}\sim 10^{-2}$ M$_{\odot}$ yr$^{-1}$ can explain the overall luminosity of the bolometric LC. 
Here, the best-fit model predicts the velocity of the interaction shock as $\sim 6000$ km s$^{-1}$, while the narrow allowed lines in the nebular spectra (\ion{O}{i,} \ion{Ca}{ii}, \ion{Mg}{ii}, and \ion{He}{I}; see Fig.~\ref{fig_lineprof}), which should originate from the gas in the shock shell, and not from the unshocked ejecta, indicate that some interaction regions of the SN ejecta were decelerated to $\sim 1000$ km s$^{-1}$. This discrepancy suggests that the actual CSM must be confined to smaller regions and thus denser than the spherical CSM assumed in the LC modeling. 
At the same time, the coexistence of the narrow and broad components of the ejecta emission lines in the nebular spectra of SN~2021efd further supports a clumpy distribution for the CSM rather than a distribution with multiple rings of different sizes on a single orbital plane. This spectral feature requires multiple emitting regions at different locations of the ejecta, while, in the multiple-ring scenario, the interaction -  and thus the energy input - occurs at a specific radius at a given time. A schematic representation of the inferred CSM geometry is shown in Fig.~\ref{schematic}.

\section{Discussion} \label{sec:discussion}

\subsection{Progenitor system of SN~2021efd}

In this subsection, we discuss the potential progenitor of SN~2021efd. As shown in Sect.~\ref{sec:SN_properties}, the estimated ejecta properties of SN~2021efd (the velocity, mass, kinetic energy, and Ni mass) are within the diversity of typical Type~Ib and Ic SNe. This might suggest that the progenitor star of SN~2021efd is similar to those of typical Type~Ib and Ic SNe, whose ZAMS masses are likely $\lesssim 20$ M$_{\odot}$ \citep[e.g.,][]{bersten14,yoon17, dessart24}.

\begin{figure}[t]
    \centering
    \includegraphics[width=0.4\textwidth]{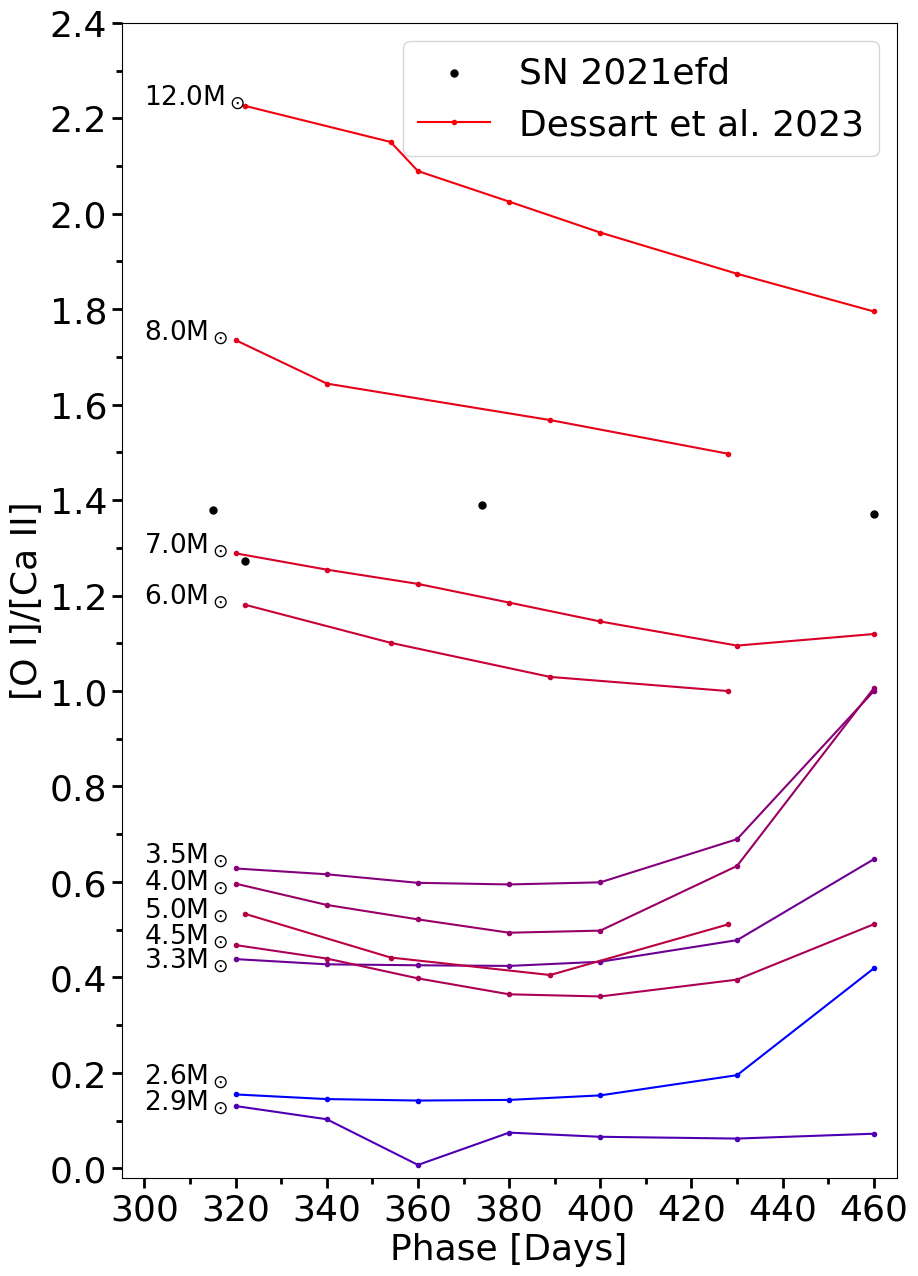}
    \caption{[[\ion{O}{i}]/[\ion{Ca}{ii}] ratio of models with different He star masses measured from synthetic spectra presented in \cite{dessart23}. The different lines correspond to different He star masses of the models. The black dots are the values measured from the nebular spectra of SN 2021efd (see Sect. \ref{sec:OI/Ca}).}
    \label{oicaii}
\end{figure}

Another hint for estimating the properties of the progenitor is the [\ion{O}{i}]/[\ion{Ca}{ii}] ratio in the nebular spectra, which is often used as a measure of the ZAMS core masses of the SN progenitors.
We compare the [\ion{O}{i}] to [\ion{Ca}{ii}] ratios of SN~2021efd, which are estimated to be $\sim1.3$ from the spectra between phases from 324 days to 476 days and $\sim1.7$ from the spectra at 641 days and 771 days (see Sect.~\ref{sec:OI/Ca}), to those from the simulated spectra of SESNe in \citet{dessart23}. These spectra are calculated with different initial conditions, including the mass of the He star. We measure the [\ion{O}{i}] to [\ion{Ca}{ii}] ratios of these synthesised spectra as for the nebular spectrum of SN~2021efd. The results are plotted in Fig.~\ref{oicaii}. The [\ion{O}{i}] to [\ion{Ca}{ii}] ratio of SN~2021efd is closest to that for the model with He star mass of 7 $M_\odot$, which corresponds to a ZAMS mass of $\sim 25~ M_{\odot}$ \citep{dessart23, woosley19, ertl20}. 
Here, we note that in SN~2021efd, the dominant energy input at late phases is not the $^{56}$Co decay in the inner ejecta, but rather the CSM interaction in the outer ejecta, which might affect the [\ion{O}{i}]/[\ion{Ca}{ii}] ratio. Therefore, this progenitor mass measurement should be considered as a reference.
Although this estimate is not necessarily correct, the estimated ZAMS mass of $\sim25~M~_\odot$ for SN~2021efd is roughly comparable to those for typical Type Ib and Ic SNe. The progenitors of Type Ic SNe are typically expected to have slightly greater ZAMS masses than those of Type~Ib SNe. Depending on metallicity, a ZAMS mass of $\sim25~M_\odot$ is compatible with progenitors of both Type~Ib and Type~Ic SNe based on, for example, theoretical predictions from stellar evolution models  \citep{georgy09}, and population studies of SN environments \citep{kuncarayakti18b}.


\subsection{Mass-loss properties} \label{sec:mass-loss_properties}

The CSM properties inferred in Sect.~\ref{sec:origin} suggest that the final mass-loss processes of the progenitor of SN~2021efd before the explosion were accompanied by episodic mass ejections with mass-loss rates of $\dot{M} \lesssim 10^{-2}$ M$_{\odot}$~yr$^{-1}$.
The velocity of the CSM is uncertain. Theoretically, different mechanisms would predict different CSM velocities. The minimum value of $\sim 100$~km~s$^{-1}$ might be achieved in cases where the mass loss happens at a large distance from the progenitor star, such as the mass loss via the Roche-lobe overflow in a binary system, while the maximum value of $\sim 2000$ km~s$^{-1}$ can be realized in cases such as mass ejections from the surface of H-poor massive stars \citep[e.g.,][]{crowther07}. Here, we have an observational hint for the velocity from the nebular spectra. As discussed in Sect.~\ref{sec_CSM_geometry}, the narrow lines from allowed transitions in the nebular spectra (\ion{O}{i}, \ion{Ca}{ii}, \ion{Mg}{ii}, and \ion{S}{ii}), which have widths of $\sim 1000$~km~s$^{-1}$, should originate from the gas in the shock shell. This indicates that some interaction regions should be decelerated to $\sim 1000$~km~s$^{-1}$, and thus that the CSM velocity should be lower than this velocity. Therefore, we adopt 100 and 1000~km~s$^{-1}$ as the minimum and maximum values of the mass-loss velocity, respectively. In the following discussions, we use the minimum value of $v_\mathrm{csm}=100~\mathrm{km}~\mathrm{s}^{-1}$, which produces the minimum value of the necessary mass-loss rate, as a fiducial value, and also discuss the case with the maximum value assuming $v_\mathrm{csm}=1000~\mathrm{km}~\mathrm{s}^{-1}$.

The timescale of the mass ejections is estimated from the distances of the CSM clumps to the progenitor star. For simplicity, a characteristic velocity of 10,000 km s$^{-1}$ is adopted for the outer ejecta. Since the peaks of the first and second rebrightenings are at $\sim 90$ and $\sim 120$~days after the explosion, respectively, we can estimate the distances of the CSM clumps that are responsible for these rebrightenings to be $8\times 10^{15}$ and $1\times 10^{16}$ cm. Such CSM clumps would have been ejected at $\sim 25$ and $\sim 32$ years before the SN explosion, assuming the CSM velocity of 100~km~s$^{-1}$. If the CSM velocity is 1000~km~s$^{-1}$, these values would instead be $\sim 2.5$ and $\sim 3.2$~years before the explosion. The duration of $\sim 30$~days for these rebrightenings corresponds to the time duration of the mass ejections to be $\sim 0.8-8$~years, depending on the assumed CSM velocity.
The first and second rebrightenings correspond to mass-loss rates of approximately $\sim 5\times 10^{-3}- \sim 10^{-2}~M_\odot~\mathrm{yr}^{-1}$ with durations of about 8 years, assuming a CSM velocity of 100km s$^{-1}$. In contrast, for a CSM velocity of 1000~km~s$^{-1}$, the corresponding mass-loss rates are higher by a factor of ten, while the durations are shorter by a factor of ten. Therefore, regardless of the assumed CSM velocity, the masses of these CSM clumps are estimated to be about $0.04 - 0.2$ M$_{\odot}$.
The CSM interaction continues at least until $\sim 770$ days after the explosion, which is the epoch of the last photometric detection. Thus, the extensive mass loss would have lasted at least from $\sim 220$ to $\sim 12$ years before the explosion with the assumption of $v_{\rm{csm}}=100$ km s$^{-1}$ (from $\sim 22$ to $\sim 1.2$ years with the assumption of $v_{\rm{csm}}=1000$ km s$^{-1}$).
It is likely that this extensive mass loss would have been ongoing even before this period, simply because our last photometric observation does not suggest a decline. It might also be possible that there were minor mass ejections even after this period until the explosion, whose signs would have been hidden by the SN component of SN~2021efd.

\subsection{Mass-loss mechanism}

In this subsection, we discuss possible mass-loss mechanisms to create the CSM estimated for SN~2021efd based on its derived properties. Combining constraints on the progenitor and CSM properties, the progenitor of SN~2021efd was likely a He star, similar to those of typical Type Ib SNe, originating from stars with ZAMS masses of $\lesssim 25$ M$_{\odot}$ \citep[e.g.,][]{yoon17,dessart24}. In addition, the progenitor system should have a mass loss rate of 10$^{-2}~M_\odot~\mathrm{yr}^{-1}$ at least from $\sim 130$ to $\sim 25$ years before the explosion with the assumption of $v_{\rm{csm}}=100$ km s$^{-1}$ ($0.1~M_\odot~\mathrm{yr}^{-1}$ from $\sim 13$ to $\sim 2.5$ years with the assumption of $v_{\rm{csm}}=1000$ km s$^{-1}$), which is much larger than the typical mass loss from massive stars including He stars \citep[e.g.,][]{smith2014}. This implies some extensive unknown type of mass ejections in a similar He star to the progenitors of typical Type Ib SNe, which we have rarely observed. 
If this extensive mass loss had lasted for a few thousands of years before the explosion in the scenario with $v_{\rm{csm}}=100$ km s$^{-1}$ (a few hundred years for the scenario with $v_{\rm{csm}}=1000$ km s$^{-1}$), the progenitor star would have become a carbon-oxygen-rich massive star, which would have exploded as a Type Ic SN. Given that the properties of SN~2021efd are similar to those of typical Type~Ib and Type~Ic SNe, the progenitor of SN~2021efd might have been a He star that was indeed on the way to becoming a potential progenitor of a Type~Ic SN, but ended up as a Type Ib SN because of the timing of the explosion or late initiation of the mass loss. Thus, it is important to study the mass-loss mechanism in objects like SN~2021efd for a better understanding of the He-layer stripping in the progenitors of Type~Ic SNe.

The first thing that we can say about the mass-loss mechanism is that the CSM appears to originate from the progenitor star. Since the CSM should be H-poor gas, it is natural to consider that the CSM gas should originate from the He star progenitor rather than from the potential companion star. If we consider the companion star as the origin of the CSM, it should be an H-poor star. In this case, since the main mechanism for the H-layer stripping is likely due to binary interaction \citep[e.g.,][]{yoon17}, the original H-rich gas of the companion star should transfer into the primary star, and the primary star should become a H-rich star. However, this is contradicted by the fact that the primary star exploded as a Type~Ib SN. Thus, the H-poor CSM should originate from the He star progenitor that exploded as the Type~Ib SN~2021efd.

One possible mechanism to remove the outer parts of the progenitor might be binary interaction, where the primary He star fills the Roche potential and transfers its gas into the companion star \citep[e.g.,][]{podsiadlowski92,yoon10}. However, the expected properties of the CSM in the binary interaction scenario do not match those inferred for SN~2021efd. The expected mass-loss rate in the binary scenario ($\lesssim 10^{-4}$ M$_{\odot}$~yr$^{-1}$; \citealt{ercolino25}) is much smaller than that estimated for SN~2021efd ($\gtrsim 10^{-2}$ M$_{\odot}$~yr$^{-1}$; see Sect.~\ref{sec:origin}). In addition, it is difficult to create not only episodic mass ejections with a period of $\lesssim 10$ years \citep[e.g.][]{ouchi17} but also clumpy mass ejections accompanied by the CSM clumps \citep[e.g.,][]{vetter24}. Therefore, the He layer stripping is probably due to some mass ejections caused by the He star itself.

The mass ejections might be triggered by some instability related to the last burning phases in the massive-star evolution, which might be similar to the phenomenon that causes the confined CSM observed in Type~II SNe \citep[e.g.,][]{khazov16,yaron17,boian20,bruch21}. For this scenario, the rarity of the SN~2021efd-like events, that is, Type~Ib SNe interacting with massive clumpy CSM (so far SNe 2019tsf and 2021efd), might be problematic. In addition, the CSM mass estimated for SN~2021efd is close to the upper end of those of the confined CSM of Type~II SNe \citep[$\lesssim1$ M$_{\odot}$; e.g.,][]{Forster2018}, and its extension is larger than the confined CSM \citep[$\lesssim 10^{15}$ cm; e.g.,][]{yaron17}, although it is more difficult to remove the mass from a H-poor star than the H-rich progenitors of Type~II SNe. It might be natural to interpret that the progenitor of SN~2021efd was in an unstable state caused by another reason.

SN~2021efd might be related to other types of SNe. 
If the active mass-loss episode observed in SN~2021efd had occurred much earlier, the SN could have appeared as a regular Type Ib SN or Type Ic SN, depending on how much He remained at the surface. Conversely, if the mass-loss episode had started slightly earlier and lasted longer until the explosion, the SN might appear as a Type Ic SN interacting with H-poor CSM.
In fact, some Type~Ic SNe interacting with massive H-poor CSM show similar observational properties (e.g., SN~2010mb; \citealt{ben-ami14}, SN~2021ocs; \citealt{kuncarayakti22}, SN~2022xxf; \citealt{kuncarayakti23}). These transients might be related to each other. The clumpy distribution and high mass-loss rate of the CSM in SN~2021efd resemble the giant eruptions of luminous blue variables (LBVs), although they are considered to be H-rich stars \citep[e.g.,][]{Smith2011}. The mass-loss mechanism in SN~2021efd might be related to that for LBVs, although it is also under debate \citep[e.g.,][]{Smith2011}.

Another speculative origin might be a merged He-rich star that has a peculiar ratio between the core and the envelope, creating an unstable state. Depending on the parameters of the original merging stars and the timing of the merger to the explosion, such merged stars might explode as Type~IIL or Type~IIn SNe (explosions of LBVs), normal Type~Ib or Type~Ic SNe, SN~2021efd-like events, or Type~Ic-CSM SNe.

SN~2019tsf has a lot of similarities with SN~2021efd. When it appeared from solar conjunction, it had a similar $r$-band absolute magnitude to SN~2021efd at peak. Unlike SN~2021efd, it does not show a decay rate similar to regular SESNe at first, which may mean that it is already at a later epoch, or that the CSM was in closer proximity to the progenitor. In the multiband LC presented in Fig.~2 of \citet{zenati22}, it has at least two additional bumps after the first detection, which was already after the first peak. The spectrum at the nebular phase is similar and shows the excess at the blue end like SN~2021efd, but no narrow lines are present in SN 2019tsf. Overall, we consider that SN~2019tsf might come from a similar progenitor system as SN~2021efd. One potential avenue to understanding the mass-loss histories of systems like these might be radio and X-ray observations.

\section{Conclusion} \label{sec:conclusion}

We have performed an in-depth analysis of the peculiar Type~Ib SN~2021efd. The photometric and spectroscopic properties of this object are typical for a Type~Ib SN until $\sim$30 days after the epoch of maximum light. After this epoch, the LC of SN~2021efd begins to exhibit an excess of luminosity until the end of the LC at $\sim 770$ days, with at least two clear peaks at $\sim75$ and $\sim 105$ days from the maximum light. The early spectra of SN~2021efd are not significantly different from those of typical Type~Ib SNe. The velocities of \ion{Fe}{ii} $\lambda5169$ and \ion{He}{i} $\lambda5876$ in the early phase spectra are consistent with those seen in other SESNe. The nebular phase ($\gtrsim 300$ days) spectra, however, contain features common in interacting SNe. There is an excess in the blue continuum, and almost all of the lines in the nebular spectra have a narrow component and a broad component. The narrow components are $\sim1000$~km~s$^{-1}$ wide, probably originating from the shocked regions (allowed lines) and the SN ejecta around the shocked regions (forbidden lines). No signatures of hydrogen are detected at any epoch.
The observational features of SN~2021efd resemble those of SN~2019tsf, suggesting comparable progenitor systems.

The derived ejecta and CSM properties are as follows: SN~2021efd has an ejecta mass of $M_\mathrm{ej}=2.2 \pm 0.3~M_\odot$, a kinetic energy of $E_\mathrm{k}=(0.91 \pm 0.12)\times 10^{51}$ erg, and a $^{56}$Ni mass of $M_\mathrm{Ni}=0.14 \pm 0.01~M_\odot$. The estimated progenitor mass of SN~2021efd from the [\ion{O}{i}]/[\ion{Ca}{ii}] ratio is $\sim 25~M_{\odot}$, although this value is uncertain due to the omission of the effects of CSI on the ratio. As for the CSM features, the estimated mass-loss rate of the progenitor is roughly $\sim 5\times 10^{-3}-10^{-2}~M_\odot~\mathrm{yr}^{-1}$ for the last few decades ($\sim 8-32$ years, depending on the assumed CSM velocity) and slightly smaller before this final phase. The minimum CSM mass estimated by the photometric observations is a few times $10^{-1}~M_\odot$. The observational properties suggest a clumpy CSM distribution.

The derived CSM properties have important information on the He-layer stripping mechanism of Type Ic SNe progenitors. If such extensive mass loss had continued for a longer time than the actual case before the explosion, the progenitor of SN 2021efd could have lost more helium and become a carbon-oxygen-rich star, exploding as a Type~Ic SN. Since the ejecta of SN~2021efd have properties similar to both Type Ib and Ic SNe, its progenitor was likely a He star in transition toward becoming a Type Ic SN. Studying this mass-loss process using our derived CSM properties gives a hint at understanding how He-layers are stripped in progenitors of Type Ic SNe.

\begin{acknowledgements}

We thank Qiliang Fang, Anders Jerkstrand, and Stan Barmentloo for helpful discussions.
NP, TN, and HK acknowledge support via the Research Council of Finland (grant 340613).
H.K. and T.N. were funded by the Research Council of Finland projects 324504, 328898, and 353019.
M.D. Stritzinger and S. Bose are funded by the Independent Research Fund Denmark (IRFD, grant number  10.46540/2032-00022B) and by an Aarhus University Research Foundation Nova project (AUFF-E-2023-9-28).
 T.K. acknowledges support from the Research Council of Finland project 360274.
Based on observations made with the Nordic Optical Telescope (NOT), owned in collaboration by the University of Turku and Aarhus University, and operated jointly by Aarhus University, the University of Turku and the University of Oslo, representing Denmark, Finland and Norway, the University of Iceland and Stockholm University at the Observatorio del Roque de los Muchachos, La Palma, Spain, of the Instituto de Astrofisica de Canarias. The NOT data were obtained under program ID P62-505." 
Based on observations collected at the European Organisation for Astronomical Research in the Southern Hemisphere under ESO programs 105.20DF.001 and 108.2282.001 (PI: Kuncarayakti).
This work has made use of data from the Asteroid Terrestrial-impact Last Alert System (ATLAS) project. The Asteroid Terrestrial-impact Last Alert System (ATLAS) project is primarily funded to search for near earth asteroids through NASA grants NN12AR55G, 80NSSC18K0284, and 80NSSC18K1575; byproducts of the NEO search include images and catalogs from the survey area. This work was partially funded by Kepler/K2 grant J1944/80NSSC19K0112 and HST GO-15889, and STFC grants ST/T000198/1 and ST/S006109/1. The ATLAS science products have been made possible through the contributions of the University of Hawaii Institute for Astronomy, the Queen’s University Belfast, the Space Telescope Science Institute, the South African Astronomical Observatory, and The Millennium Institute of Astrophysics (MAS), Chile.
Based on observations obtained with the Samuel Oschin Telescope 48-inch and the 60-inch Telescope at the Palomar Observatory as part of the Zwicky Transient Facility project. ZTF is supported by the National Science Foundation under Grant No. AST-2034437 and a collaboration including Caltech, IPAC, the Weizmann Institute of Science, the Oskar Klein Center at Stockholm University, the University of Maryland, Deutsches Elektronen-Synchrotron and Humboldt University, the TANGO Consortium of Taiwan, the University of Wisconsin at Milwaukee, Trinity College Dublin, Lawrence Livermore National Laboratories, IN2P3, University of Warwick, Ruhr University Bochum, Cornell University, and Northwestern University. Operations are conducted by COO, IPAC, and UW.
 The ZTF forced-photometry service was funded under the Heising-Simons Foundation grant \#12540303 (PI: Graham). The Gordon and Betty Moore Foundation, through both the Data-Driven Investigator Program and a dedicated grant, provided critical funding for SkyPortal.  
SED Machine is based upon work supported by the National Science Foundation under Grant No. 1106171 
Zwicky Transient Facility access was supported by Northwestern University and the Center for Interdisciplinary Exploration and Research in Astrophysics (CIERA).
This research has made use of the Spanish Virtual Observatory (https://svo.cab.inta-csic.es) project funded by MCIN/AEI/10.13039/501100011033/ through grant PID2020-112949GB-I00.
IS is supported by the PRIN-INAF 2022 project “Shedding light on the nature of gap transients: from the observations to the models”.
AR acknowledges financial support from the GRAWITA Large Program Grant (PI P. D'Avanzo) and from the PRIN-INAF 2022 "Shedding light on the nature of gap transients: from the observations to the models".
K.M. acknowledges support from the Japan Society for the Promotion of
Science (JSPS) KAKENHI grant (JP24KK0070, JP24H01810). The work is partly
supported by the JSPS Open Partnership Bilateral Joint Research Projects
between Japan and Finland (K.M and H.K; JPJSBP120229923).
S.M. acknowledges financial support from the Research Council of Finland project 350458.
L.G. acknowledges financial support from AGAUR, CSIC, MCIN and AEI 10.13039/501100011033 under projects PID2023-151307NB-I00, PIE 20215AT016, CEX2020-001058-M, ILINK23001, COOPB2304, and 2021-SGR-01270.
AGY’s research is supported by ISF, IMOS and BSF grants, as well as the André Deloro Institute for Space and Optics Research, the Center for Experimental Physics, a WIS-MIT Sagol grant, the Norman E Alexander Family M Foundation ULTRASAT Data Center Fund, and Yeda-Sela;  AGY is the incumbent of the The Arlyn Imberman Professorial Chair.
This material is based upon work supported by the National Science Foundation Graduate Research Fellowship Program under Grant Nos. 1842402 and 2236415. Any opinions, findings, conclusions, or recommendations expressed in this material are those of the author(s) and do not necessarily reflect the views of the National Science Foundation.
CPG acknowledges financial support from the Secretary of Universities
and Research (Government of Catalonia) and by the Horizon 2020 Research
and Innovation Programme of the European Union under the Marie
Sk\l{}odowska-Curie and the Beatriu de Pin\'os 2021 BP 00168 programme,
from the Spanish Ministerio de Ciencia e Innovaci\'on (MCIN) and the
Agencia Estatal de Investigaci\'on (AEI) 10.13039/501100011033 under the
PID2023-151307NB-I00 SNNEXT project, from Centro Superior de
Investigaciones Cientificas (CSIC) under the PIE project 20215AT016 and the program Unidad de Excelencia Maria de Maeztu CEX2020-001058-M, and from the Departament de Recerca i Universitats de la Generalitat de
Catalunya through the 2021-SGR-01270 grant.

\end{acknowledgements}

\bibliographystyle{aa} 

\bibliography{export.bib}

\end{document}